
\input harvmac.tex
%
%
%
%

\hyphenation{anom-aly anom-alies coun-ter-term coun-ter-terms
dif-feo-mor-phism dif-fer-en-tial super-dif-fer-en-tial dif-fer-en-tials
super-dif-fer-en-tials reparam-etrize param-etrize reparam-etriza-tion}

%
%
\newwrite\tocfile\global\newcount\tocno\global\tocno=1
\ifx\bigans\answ \def\tocline#1{\hbox to 320pt{\hbox to 45pt{}#1}}
\else\def\tocline#1{\line{#1}}\fi
\def\toclead{\leaders\hbox to 1em{\hss.\hss}\hfill}
\def\tnewsec#1#2{\xdef #1{\the\secno}\newsec{#2}
\ifnum\tocno=1\immediate\openout\tocfile=toc.tmp\fi\global\advance\tocno
by1
{\let\the=0\edef\next{\write\tocfile{\medskip\tocline{\secsym\ #2\toclead\the
\count0}\smallskip}}\next}
}
\def\tnewsubsec#1#2{\xdef #1{\the\secno.\the\subsecno}\subsec{#2}
\ifnum\tocno=1\immediate\openout\tocfile=toc.tmp\fi\global\advance\tocno
by1
{\let\the=0\edef\next{\write\tocfile{\tocline{ \ \secsym\subsecsym\
#2\toclead\the\count0}}}\next}
}
\def\tappendix#1#2#3{\xdef #1{#2.}\appendix{#2}{#3}
\ifnum\tocno=1\immediate\openout\tocfile=toc.tmp\fi\global\advance\tocno
by1
{\let\the=0\edef\next{\write\tocfile{\tocline{ \ #2.
#3\toclead\the\count0}}}\next}
}
%
%

%
%
%
%

%
\long\def\optional#1{}
\def\cmp#1{#1}         
%
%
\let\narrowequiv=\equiv
\def\equiv{\;\narrowequiv\;}

\def\tilde{\widetilde}
\fontdimen16\tensy=2.7pt\fontdimen17\tensy=2.7pt 



%

\def\th{\theta}
\def\ph{\varphi}

\def\ket#1{| #1 \rangle}         
\def\bra#1{\langle #1 |}         
\def\pa{\partial}
%
\def\dd{\mskip 1.3mu{\rm d}\mskip .7mu} 
%

\global\newcount\figno \global\figno=1
\newwrite\ffile
\def\pfig#1#2{Fig.~\the\figno\pnfig#1{#2}}
\def\pnfig#1#2{\xdef#1{Fig. \the\figno}%
\ifnum\figno=1\immediate\openout\ffile=figs.tmp\fi%
\immediate\write\ffile{\noexpand\item{\noexpand#1\ }#2}%
\global\advance\figno by1}

%
%
\def\tfig#1{Fig.~\the\figno\xdef#1{Fig.~\the\figno}\global\advance\figno by1}

\def\insfig#1#2#3#4#5#6#7#8#9{\midinsert\vskip #3truein
\includegraphics{#4}
    \narrower\narrower\noindent#1:#2\endinsert}
%
%
\def\pa{\partial}
\def\dbd#1{{\del \over \pa{#1}}}
\def\der#1#2{{{\pa{#1}} \over {\pa{#2}}}}

\def\rr{{\tilde r}}
\def\rb{\overline{\vphantom\i {r}}}
\def\tb{\overline{\vphantom\i {t}}}
\def\rrb{\overline{\rr}}
\def\rrho{\tilde \rho}
\def\aa{{\tilde a}}

\def\y{(r-\rr)}
\def\ay{|r-\rr|}
\def\fp{f'(\ay)}
\def\gp{g'(|r|)}
\def\gpy{g'(|y-\rr|)}
\def\hp{h'(|\rr|)}
\def\hpy{h'(|y+r|)}
\def\mh#1#2{\widehat {\cal M}_{#1,#2}}
\def\ph#1#2{\widehat {\cal P}_{#1,#2}}
\def\mgs#1#2{{\cal M}_{#1,#2}}
\def\pgs#1#2{ {\cal P}_{#1,#2}}

\def\sc{\check \sigma}
\def\s{\sigma}
\def\sth{\sc}
\def\g{\gamma}
\def\th{\theta}
\def\dilone{-{1 \over 2} c_1 \g _{-1/2} \delta (\g _{1/2})\ket{0}}
\def\diltwo{-2 c_1 \bar c_1 \bar c_{-1} \beta _{-1/2} \delta (\g _{1/2})
                \ket{0}}
\def\dilonep#1{c^{(#1)}_1 {\g}^{(#1)}_{-1/2} 
 \delta ({\g}^{(#1)}_{1/2})\ket{0}^{#1}}
\def\diltwop#1{c^{(#1)}_1 \bar c^{(#1)}_1 \bar c^{(#1)}_{-1}
 \beta^{(#1)}_{-1/2} \delta (\g^{(#1)}_{1/2}) \ket{0}^{#1}} 

\def\Done{\ket{D_1}}
\def\Dtwo{\ket{D_2}}

\def\b#1#2{{b}^{(#2)}_{#1}}
\def\bb#1#2{{\bar b}^{(#2)}_{#1}}
\def\bbc#1{{\bar b}_{#1}}
\def\bp#1{b[\s_*({\dbd #1})]}
\def\sps{\ket{psi}}

\def\we{\wedge}
\def\vf{\dd \rr\we \dd \rrb\we \dd r\we\dd \rb}
\def\tf{\dd r\we \dd \rb}
\def\tff{\dd \rr\we \dd \rrb}
\def\tfa{\dd \rr\we \dd r\we\dd \rb}
\def\tfb{\dd \rr\we \dd \rrb\we \dd r}
\def\yfa{\dd y\we \dd r\we\dd \rb}
\def\yfb{\dd \rr\we \dd \rrb\we \dd y}
\def\rmeas{[\dd r\we\dd\rb|\dd\rho]}
\def\tmeas{[\dd t\we\dd\tb|\dd\tau]}
\def\nsub{(r^2 (1-g)+g (y+r)^2)}
\def\msub{(r+g y)}
\def\Ot{\tilde \Omega}
\def\Oht{\widehat{\tilde \Omega}}
\def\Om{\Omega}

\def\vt#1{\tilde V_{#1}}
\def\Ut#1{\tilde \Upsilon_{#1}}

\def\svt#1{\s_*(\tilde V_{#1})}
\def\db#1{\delta[\beta_{#1}]}
\def\dg#1{\delta[{#1}]}
\def\dbp#1{\delta'[\beta_{#1}]}

\def\dbt#1#2{\delta[\beta^{(#2)}_{#1}]}
\def\dbpt#1#2{\delta'[\beta^{(#2)}_{#1}]}
\def\dbppt#1#2{\delta''[\beta^{(#2)}_{#1}]}
\def\bv#1{b[v_{#1}]}
\def\Bv#1{B[v_{#1}]}
\def\B#1{B[\s_*({#1})]}
\def\dBnu#1{\delta[B[\nu_{#1}]]}
\def\be#1{\beta_{#1}}
\def\bet#1#2{\beta^{(#2)}_{#1}}
\def\cb#1{{\bar c}_{#1}}
\def\mha{-{1\over2}}
\def\kpkq#1#2{\ket{#1}^P\otimes\ket{#2}^Q}
\def\drrb{\dbd{\rrb}}
\def\drb{\dbd{\rb}}
\def\fpt{f'(|t|)}
\def\pp{\tilde p}
\def\bh{\hat b}
\def\qb{\bar{q}}
%
%
%
\lref\DUNP{J. Distler, unpublished.}%
\lref\DESRST{J. Distler and P. Nelson, ``The dilaton equation in
semirigid string theory," Nucl. Phys. {\bf B366} (1991) 255.}%
\lref\GOV{S. Govindarajan, P.Nelson, and E. Wong, ``Semirigid geometry",
Penn U. preprint, UPR-0477T, Sept. 1991. }%
\lref\PRIV{J. Distler and P. Nelson, private communication}%
\lref\EWTG{E. Witten, ``On
the structure of the topological phase of two-dimensional gravity,''
Nucl. Phys. {\bf B340} (1990) 281.}%
\lref\VVTG{E. Verlinde and H. Verlinde, ``A solution of two-dimensional
topological gravity,'' Nucl. Phys. {\bf B348} (1991) 457.}%
\lref\semirig{J. Distler and P. Nelson, ``Semirigid supergravity,''
Phys. Rev. Lett. {\bf66} (1991) 1955.}%
\lref\CIGVO{P. Nelson, ``Covariant insertions of general vertex operators,"
Phys. Rev. Lett. {\bf62} (1989) 993.}%
\lref\EFEFS{H.S. La and P. Nelson, ``Effective field equations for
fermionic strings," Nucl. Phys. {\bf B332} (1990) 83.}%
\lref\TCCT{J. Distler and P. Nelson, ``Topological couplings and
contact terms in 2d field theory,'' Commun. Math. Phys. {\bf138} (1991) 273.}%
\lref\JDTG{J. Distler, ``2-d quantum gravity, topological field theory,
and the multicritical matrix models,'' Nucl. Phys. {\bf B342} (1990) 523.}%
\lref\AGNSV{L. Alvarez-Gaum\'e, C. Gomez, P. Nelson, G. Sierra, and C.
Vafa, ``Fermionic strings in the operator formalism," Nucl. Phys.
{\bf B311} (1988) 333.}%
\lref\AGETAL{L. Alvarez-Gaum\'e, C. Gomez, G. Moore, and C.
Vafa, ``Strings in the operator formalism," Nucl. Phys.
{\bf B303} (1988) 455.}%
\lref\JPvert{J. Polchinski, ``Vertex operators in the Polyakov path
integral,'' Nucl. Phys. {\bf B289} (1987) 465.}%
\lref\LSMS{P. Nelson, \cmp{``Lectures on strings and moduli space,''} Phys.
Reports {\bf149} (1987) 304.}%
\lref\EUGENE{E. Wong, ``Recursion relations in semirigid topological
gravity'', Penn U. preprint UPR-0491T, Nov. 1991.}%
\Title{PUPT-1296}
{\vbox{\centerline{Dilaton Contact Terms in the}
	\vskip2pt\centerline{Bosonic and Heterotic Strings}}}
\centerline{Mark D. Doyle\footnote{$^\dagger$}
{mdd@puhep1.princeton.edu}\footnote{$^\ddagger$}
{Research supported in part by NSF Grant PHY90-21984}}
\bigskip\centerline{Joseph Henry Laboratories}
\centerline{Jadwin Hall}
\centerline{Princeton University}\centerline{Princeton, NJ 08544 USA}

\vskip .3in
Dilaton contact terms in the bosonic and heterotic strings are examined
following the recent work of Distler and Nelson on the bosonic
and semirigid strings.  In the bosonic case dilaton two-point functions
on the sphere are calculated
as a stepping stone to constructing a `good' coordinate family for
dilaton calculations on higher genus surfaces. It is found that
dilaton-dilaton contact terms are improperly normalized, suggesting that
the interpretation of the dilaton as the first variation of string coupling
breaks down when other dilatons are present. It seems likely that this
can be attributed to the tachyon divergence found in \TCCT.
For the heterotic case, it is found that there is no tachyon divergence
and that the dilaton contact terms are properly normalized.
Thus, a dilaton equation analogous to the one in topological gravity
is derived and the interpretation of the dilaton as the string
coupling constant goes through.

\Date{01/92} 

\newsec{Introduction}
Recently there has been much progress towards a non-perturbative
definition of string theory with the introduction of the matrix
models and topological field theories. While these two
approaches seem vastly dissimilar on the surface, it has become
abundantly clear over the past year or so that there are deep
connections between them. It is also clear that some of the deeper
insights into the nature of string theory are given in terms of the
geometry of Riemann surfaces. In fact, many of the surprising features
of string theory are found to have a simple, concise, and natural
explanations when phrased in geometrical terms. Similarly, a geometric
approach to topological gravity based on $N=2$ semirigid geometry
\semirig \DESRST \GOV \EUGENE has shed
light on the nature of the contact interactions that give rise to
the correlation functions and the recursion relations in topological
gravity. The primary focus of this recent work has been to establish the
`puncture' and `dilaton equations' of \EWTG and \VVTG in this
geometrical framework.  This approach is modeled
on the dilaton contact terms in the bosonic string\TCCT. Here the
`dilaton equation' is the well-known low-energy theorem  that the
zero-momentum dilaton couples to the string coupling constant.
The reason for this
is that the (zero-momentum) dilaton in the bosonic string and all
of the operators in topological gravity (except the puncture operator)
are BRST-exact and na\"\i vely
decouple\JDTG. But, as pointed out in \CIGVO, we must be careful.
We are concerned here with the `equivariant', or relative,
BRST-cohomology and while the states we
are considering are BRST-trivial, they are not trivial in the
equivariant cohomology. In the full (absolute) cohomology, states are
trivial if
they can be written as the BRST-operator $Q$ acting on another state.
However, we are really interested in states that obey an equivariance
condition $$(b_0-\bar{b}_0)\ket{\psi}=0.$$ In the equivariant cohomology
states satisfying the equivariance condition are in the same class if
they differ by $Q$ acting on a state that also obeys the equivariance
condition. Thus a state that is trivial in the full cohomology (i.e., it
can be written as $Q$ acting on another state) may not be trivial
in the equivariant cohomology if
the state it is $Q$ of does not satisfy the equivariance condition.
This is precisely the case of the dilaton in critical string theory and
the operators in topological gravity. Below we will see how the failure of
this equivariance condition prevents states like the dilaton from
decoupling.

A careful analysis of the geometry pertinent to the contact interactions
in the bosonic and semirigid strings has been carried out in \TCCT and
\DESRST. In both cases, the analysis was carried out in an operator
formalism \AGETAL \CIGVO \EFEFS \semirig, in which
correlation functions are calculated by associating each amplitude to a
punctured (Semirigid Super-) Riemann surface with operators inserted at
the punctures.
Normal ordering requires the introduction of a coordinate at the
puncture. This is particularly important for the dilaton and the other
operators of topological gravity, because it is the ghost insertions due
to the normal-ordering prescription that actually give non-zero
answers. The calculations of \TCCT and \DESRST depended implicitly on the
existence of what might be called a `good' coordinate family.

Imagine
inserting an operator onto a Riemann surface. To do so requires the
introduction of a coordinate family that is appropriate for the entire
moduli space of the
surface, as the point moves around. In addition, while any coordinate
family would give equivalent integrated answers, we would like to make
our coordinate family as convenient as possible. In particular, we are
concerned with integrating over the location of one of the punctures
(the one where the dilaton is inserted)
while holding everything else fixed. That is, we would like to integrate
over the position of the dilaton while keeping the moduli of the
surface and the other punctures fixed. In \TCCT, the properties of a
convenient coordinate family for establishing the dilaton equation
were outlined. Firstly, we should take advantage of simplicity offered
by using a holomorphic coordinate family wherever possible. It would be
wonderful if we could use an entirely holomorphic family; however, such
global holomorphic family of coordinates do not exist in general.
Furthermore, it would easier to integrate over the location of the
puncture holding the other moduli fixed, if we use coordinates for the
puncture that are independent of the moduli for locations of the other
punctures and for the surface itself. This can be done in regions of
moduli space where the puncture is far from the others. However, this
property cannot be maintained when two punctures approach each other and
the coordinates for one puncture will depend on the location
of the other puncture \TCCT. In such a region of moduli space,
the coordinates will be given by a plumbing fixture construction in
which a `standard' universal three-punctured sphere with coordinates
is sewn to the rest of the surface. The colliding of two punctures is
then replaced by the two punctures being located on a sphere that is
pinching off from the rest of the surface. The trick is to interpolate
smoothly between these two desirable coordinate systems. This
interpolation gives rise to the non-analytic behavior that
ultimately gives rise to the contact interaction in this viewpoint.
The strength of this formalism is that the interpolation allows the
delta-function contact terms to be smoothed out away from the region
of moduli space where the two points actually collide.
In \TCCT and \DESRST, the dilaton contact terms were
calculated only in one patch of moduli space describing the approach of
one puncture to the fixed location of another puncture. All of the other
moduli and their ghost insertions were suppressed. Although it is
quite reasonable that one should be able to do this, it is not
completely clear that a global coordinate family with these properties
exists. One aim of this paper is to establish the existence of `good'
coordinate families on  general Riemann surfaces. This is done by
performing a `pants' decomposition of the surface into a set of
punctured spheres and giving a prescription for constructing a suitable
family on each sphere. Since different pants decompositions correspond to
different cells of moduli space, it still remains to show that the
families on each of the cells can be glued together continuously to give
a global family. Below we sketch how this may be accomplished.

As a stepping stone towards this goal, we will calculate two-point
functions of dilatons on the sphere in the bosonic string. This requires
the introduction of a coordinate family describing the insertion of an
operator on a three-punctured
sphere. We will see that such a family is easily given and that it is
easily generalized to give building blocks for global families for
higher genus surfaces. The dilaton-dilaton calculation on the sphere is
interesting in its own right because we will see that dilaton-dilaton
contact terms are normalized incorrectly with respect to dilaton-strong
physical state contact terms to give a correct dilaton equation. This
was first noted by Distler and Nelson \PRIV.
Thus it seems that while the insertion of a zero-momentum dilaton into a
correlation function of strong physical states (see below) behaves like
the first variation of the string coupling constant, inserting
a second dilaton is not the same as a second variation. While it seems
likely that tachyonic divergence found in \TCCT is the source of this
puzzling observation, its full significance is unclear and merits
further consideration.

To add further weight to this assertion, we go on to examine the heterotic
string where one might expect the tachyon problem to go away. Indeed, it
is found that the analogous dilaton equation goes through and that
dilaton-dilaton contact terms behave nicely. The calculation here is
done using a one patch computation following closely the work in
\TCCT and \DESRST (although it contains several new features) and makes
use of the general framework of
\EFEFS which contains a concise exposition of the operator formalism in
both the bosonic and heterotic strings and of dilatons in the heterotic
string. We hope that the reader finds this approach to be more than
satisfactory after the arguments made in the bosonic string.

In the first part of the paper we review the general framework for
examing contact
terms in the operator formalism developed in \CIGVO and\TCCT. We then
go on to re-examine some issues in the bosonic string by considering
dilaton two-point functions on the sphere where we explicitly give one
global coordinate family for the entire moduli space.
We then describe a prescription for constructing a `good' coordinate family
on higher genus surfaces by generalizing the coordinate family used on
sphere in our two-point calculations.
In the second part of the paper, we examine the dilaton contact terms in
the heterotic string in which there are new features. Before doing this
we review and develop the necessary heterotic geometry and the fermionic
operator formalism. Finally, we end with a brief discussion of some of
the significant issues raised our results.

\newsec{General Framework}
Since we will be examining several different calculations, it is perhaps
best to begin with a brief sketch of the general philosophy behind the
calculations that we have done. The general framework is that of the
operator formalism as developed in \AGETAL \AGNSV \CIGVO \EFEFS
 and the reader should look there for a more detailed exposition.

Recall that the correlation functions in bosonic string theory can be
calculated as path integrals over Riemann surfaces on which operators
have been inserted. The path integral can be reduced to an integral
of a differential form  over
the compactified moduli space of a punctured Riemann surface (the
punctures arising from the insertions of the operators). As is
well-known, the operators in the path integral are represented by vertex
operators carrying the appropriate quantum numbers. The vertex operators
require normal-ordering and it is necessary to introduce coordinates
at the punctures to do this. A particularly nice normal-ordering
prescription was given by Polchinski \JPvert.
There one uses the `flattest possible coordinate' at the puncture to
normal-order the operators and one finds that
the ghost insertions that form the measure are modified
by the normal-ordering, giving the so-called $\bh$-prescription. Most
operators in the bosonic string are unaffected by these modified
insertions, but certain `frame-dependent' operators like the dilaton
require them. The insertion of these operators is dependent on the
prescription used to normal-order them.
In fact, it is seen that the dilaton couples to the scalar
curvature of the surface precisely through these modified insertions:
The curvature gives rise to a mixing of the holomorphic and
antiholomorphic coordinates in this `flattest' coordinate system.
This can all be elegantly restated and understood through the operator
formalism in the way described below. In \CIGVO Nelson gave a beautiful
geometric interpretation of the above results and it is this treatment
that we follow.

The operator
formalism gives a general prescription for constructing the appropriate
measures on moduli space associated to particular correlation functions.
The idea is to cut disks out around each operator insertion and perform
the path integral over the rest of the surface. This information is then
represented by a state $\bra {\Sigma}$, which encodes all of the
information from the rest of the surface into a wave function on the
boundaries of the excised disks.
This is ideal for our calculations of contact terms because choosing a
`good' coordinate family (as in the introduction) allows us
to focus only on the region of interest in moduli space, namely, when the
two operators are coming close together. The rest of the surface and the
other operators are all held fixed in our state as we integrate over the
location of an operator insertion.
To be more explicit, we give the prescription for constructing the
measure in the bosonic string\CIGVO. We leave the heterotic
generalization for Section 5.

We start with a genus $g$ Riemann surface with $s$
punctures and its moduli space $\mgs gs.$ In addition, we require a
coordinate at each puncture to normal-order our insertions, and,
therefore, we consider  an infinite-dimensional augmented
moduli space of punctured Riemann surfaces with coordinates at each
puncture, denoted $\pgs gs$ . We can also use these coordinates to
excise the disks around each puncture.
$\pgs gs$ is a fiber bundle
over $\mgs gs$, where the  projection $\pi:\pgs gs \to \mgs gs$ is just
forgetting about the coordinates at the punctures.
We can construct a
measure on $\mgs gs$ from a naturally defined one on $\pgs gs$ by
pulling it back via a section $\sigma$ of $\pi:\pgs gs \to \mgs gs$.
Unfortunately, there is no global holomorphic section; however, it is
possible to find local sections that differ by $U(1)$ phases across
patch boundaries, and the resulting measure on $\mgs gs$ will be
independent of the choice of section if the states we are inserting obey
certain conditions (at least up to total derivatives).
Most physical states  in string theory are
`strong physical states (SPS)' which satisfy
 \eqn\sps{\eqalign{L_n\ket{\psi}=\bar{L}_n\ket {\psi}&=0 ,\; n \ge 0\cr
 b_n\ket{\psi}=\bar{b}_n \ket {\psi}&=0,\; n\ge 0. }}
Strong physical states have the wonderful property that the measure
constructed with them does not depend in any way on the choice of
coordinate slice.
However, this is overly restrictive, and we would like to insert many
interesting states (i.e., dilatons)  that do not obey these
conditions. In \CIGVO it is shown that states obeying a weaker set of
conditions,
\eqn\wps{\eqalign{ (L_0-\bar{L}_0)\ket{\psi}=&0\cr
                   (b_0-\bar{b}_0)\ket {\psi} =&0\;,}}
can still be inserted.
These comprise the `weak physical state conditions (WPSC)' and they are
just the condition that the measure formed with such a state be independent
of the $U(1)$ phase jumps across patch boundaries.
States obeying \wps, but not \sps are precisely the `frame-dependent'
states of Polchinski.
This has interesting consequences for the dilaton which is the primary
focus of this paper.

Since the descriptions of the measures on $\mgs gs$ and $\pgs gs$  are
more than adequately described
in other places \CIGVO \AGETAL \TCCT\EFEFS, we will only give a brief
description for completeness. Then we will specialize it to our
purposes. The measure on $\pgs gs$, in abbreviated form
\foot{We have suppressed the antiholomorphic parts and also the fact
that the vector fields $v_i$ are really $N$-tuples of vector fields.},
is
\eqn\meas{\eqalign{\Ot(\vt1,\vt2,\ldots,\vt{3g-3+s})&=\cr
  &\bra{\Sigma,z_1,\ldots}\bv1\bv2\ldots\bv{3g-3+s}
         \ket{\psi_1}_{P_1}\otimes \cdots\otimes\ket{\psi_s}_{P_s}.}}
The $\vt{i}$ are tangent vectors to $\pgs gs$ and the $v_i$ are
abstract Virasoro generators that correspond to the $\vt{i}$. The $v_i$
act on $\pgs gs$ through `Schiffer variations' and can thus be
associated with tangent vectors to $\pgs gs$. See \TCCT and \EFEFS
for details. The notation $b[v]$ corresponds to
$$\oint  b_{zz}(z) v^z(z) \dd z, $$  in which the contour integral is
performed on a contour surrounding the puncture.
To get a measure $\Omega$ on $\mgs gs$, we use the coordinate family to
pullback the measure $\Ot$,
$$\Om=\s^*\Ot\;.$$
This is given in the standard way by
\eqn\measm{\Om(V_1,V_2,\ldots,V_{3g-3+s})=
           \Ot(\svt1,\svt2,\ldots,\svt{3g-3+s}).}
Here the vectors $\vt{i}$ are any vectors that project down to the
$V_i$, the vectors tangent to $\mgs gs$. It can be shown that the
measure obtained in this way is independent of the choices made if all
of the operators are SPS. For WPS it is found that the measure will only
change by a global total derivative if the coordinate slice
differs by  $U(1)$ phases across patch boundaries, and, hence, the
integrated answers will be unaffected by the choices made. Furthermore,
it can be shown that if one
of the states $\ket{\psi}= \ket{Q\lambda}$, the measure is just
the exterior derivative, $\dd$, of
the corresponding measure formed with $\ket{\lambda}$, and, thus, that
$Q$ acts as the exterior derivative on $\pgs gs$.

In this paper we are focusing on the dilatons and their contact terms
and the above formalism is readily adapted to our purpose. It is
well-known that dilatons measure background curvature and that they couple to
the Euler characteristic of the surface. We would like to
investigate whether a `dilaton equation' similar to that of \VVTG
is valid
in the bosonic and heterotic strings. Heuristically, we have,
\eqn\dil{ \langle D O_1 O_2 \ldots O_s \rangle =
          -2 \pi i\, (2 g-2  +s)\, \langle O_1 O_2 \ldots O_s \rangle,}
where the $O_i$ maybe SPS's or other dilatons. The $(2 g-2)$ arises from
integrating the dilaton over the surface and the additional $s$
arises from contact interactions of the dilaton and the other operators.
(The factor of $-2 \pi i$ is conventional and could be absorbed by a
rescaling of the dilaton.)

In the bosonic string the zero-momentum dilaton can
be written as
\eqn\bosdil{ (Q+\bar Q) (c_0 -\bar{c}_0)\ket{0}= (c_1 c_{-1} -\bar{c}_1
       \bar{c}_{-1})\ket{0}. }
Na\"\i vely this state should
just decouple from all correlation functions since it is BRST-exact, and
the BRST-operator $Q$ corresponds to the exterior derivative on
$\pgs gs$. However, although the dilaton itself satisfies the WPSC,
$ (c_0-\bar{c}_0) \ket{0} $  is not annihilated by $(b_0-\bar{b}_0)$
and so while the measure for the dilaton is locally a total derivative,
it is sensitive to the $U(1)$ phase jumps across the patch boundaries.
Thus it is not a global total derivative and the
contributions at the patch boundaries prevents the dilaton from
decoupling;
in fact, the boundary contributions build up the Euler characteristic
of the surface \CIGVO. Closely related to the dilaton is the state
\eqn\bosdilnot{ (Q+\bar Q) (c_0 +\bar{c}_0)\ket{0}}
Here $(c_0+\bar{c}_0)\ket{0}$ is annihilated by
$(b_0-\bar{b}_0)$, and is insensitive to the $U(1)$ phases. Hence,
the state \bosdilnot does indeed give rise
to {\it global} total derivatives and it decouples from all
correlation functions. Thus, we are free to add this state to our
dilaton state and work with the new dilaton state
\eqn\bosdilhol{ \ket{D}=2 c_1 c_{-1}\ket{0}.}
Correlation functions computed with this state are the same as computed
with \bosdil because the measures formed from them differ by a genuine
global total
derivative. This purely holomorphic dilaton was used in \DESRST to avoid
the tachyonic divergence found in \TCCT which resulted from the fusion
of the holomorphic and antiholomorphic pieces in the state \bosdil.

The bosonic case was mostly investigated in \TCCT. It was found that
the contact interaction for a dilaton and an SPS depended on the
plumbing fixture coordinates (see below) used to describe the collision
of the two operators, and a physical `long thin tube'
prescription was advocated for getting the proper contact term. However,
for two dilatons, the contact contribution is $3/2$ the contact
contribution of dilaton-SPS, and the
dilaton equation does not hold for general correlation functions
containing multiple dilatons.\foot{This failure of the dilaton equation
in the bosonic string was lurking in \TCCT, but it was not made explicit
because of the focus on the tacyhon divergence there.  However, as pointed
out in \DESRST, one can work with the purely holomorphic dilaton and avoid
the divergence. In addition, another one patch calculation in \PRIV also
found this factor of 3/2 and we will later see it explicitly in the calculation
on the sphere.}
This was the main motivation
behind the construction of the semi-rigid approach to topological
gravity. There almost all of the operators are WPS's and the situation
is much
cleaner: There is no residual choice for the insertion coordinates and
the dilaton equation is always obeyed. The situation in the heterotic
string is in between the other two cases. There is no tachyon divergence
and the dilaton equation is always obeyed after making the residual
choice for the plumbing fixture coordinates.

Before presenting our calculations, we should indicate how this
formalism can be adapted to the calculation of contact terms and the
dilaton equation. The basic idea of the dilaton equation is to integrate
over the position of the dilaton and reduce the $n+1$-point function
to an $n$-point function. In the version of the calculations
developed for the semi-rigid and bosonic strings in \TCCT
\semirig \DESRST, we restrict
ourselves to the region of moduli space where the dilaton approaches one
particular operator and, working in that one patch only,  integrate only
over the moduli corresponding to the relative location of the dilaton
with respect to the other operator. This implicitly depends on the
existence of a `good' coordinate family. Below, we will give a
prescription for constructing such a family in the bosonic string.
To get a dilaton-dilaton
contact term, it is necessary to insert one of the dilatons with a
coordinate family representing a curved background. Above it was
mentioned that on a Riemann surface
curvature can be locally represented as a mixing of the holomorphic and
antiholomorphic coordinates describing the location of a point in this
one patch where the calculation is being done. This was
the method used in \TCCT and we will extend it here to the heterotic
case. For the bosonic string, we will only be working on the sphere
where the curvature is hidden in the non-analytic transition function
between the two
coordinate patches that are needed to cover the sphere.
We refer to this as the `global' picture of curvature because the
curvature is not described in a local manner. The advantage of this
description is that we can unambiguously integrate over the entire
sphere, whereas in the local calculation we are restricted to one
patch. Furthermore, we will see that this local description can interfere
with the identification of certain total derivatives  with respect to
the moduli associated to the location of the insertions.
But all is not lost and we can still extract all of the information that
we will need.

When the operators are far apart, it is appropriate
to use as coordinates on the moduli space the positions of the the
operators themselves. In this region of moduli space the coordinates at
one puncture is independent of the moduli of the others.
However, when the two operators approach each
other we are better served by making a conformal transformation to a
surface in which the two operators are pinching off from the rest of the
surface. In this region, we are forced to use coordinates for one
puncture that depend on the moduli of the other puncture.
The appropriate coordinates are
the pinching parameter $q$ associated to the plumbing fixture and the
location of where the degeneration is occurring. To specialize even more,
we will use the fixture depicted in \tfig\bplumb. In this case,
integrating over $q$ will correspond to integrating over the position of
the dilaton inserted at $P$ and will leave us with the other operator
always at the point $Q$. Our approach is then to choose a coordinate
slice that smoothly interpolates between the two pictures. Furthermore,
by choosing a `good' coordinate family,
we can keep all of the other moduli associated to the
surface and the locations of the other operators fixed as we integrate
over the location of the dilaton. As mentioned
above, the operator formalism allows us to summarize all of this in to
a state $\bra{\Sigma}$ and we will never have to explicitly display
the dependence
on the moduli not associated to the locations of the operators that we are
considering. This is a substantial simplification. As explained in \TCCT,
our interpolation will have the effect
of smoothing the delta-function contact interaction out away from $q=0$
to an annulus in the $q$-plane.

\insfig\bplumb{Bosonic plumbing fixture with coordinates. The top figure
shows coordinates that are appropriate for when $P$ and $Q$ are far
apart. The coordinates at $P$ are independent of the location of $Q$.
The lower figure depicts $P$ and $Q$ on a sphere pinched off from the
rest of the surface, appropriate for when they are close together.
Now the coordinates at $P$ explicitly depend on the
coordinates at $Q$. The plumbing fixture places $Q$ at 0 and is sewn to
the original position of $Q$ on the surface. This ensures that after
integrating over $P$ we are left with insertions at the original
location of $Q$.}{2.5}{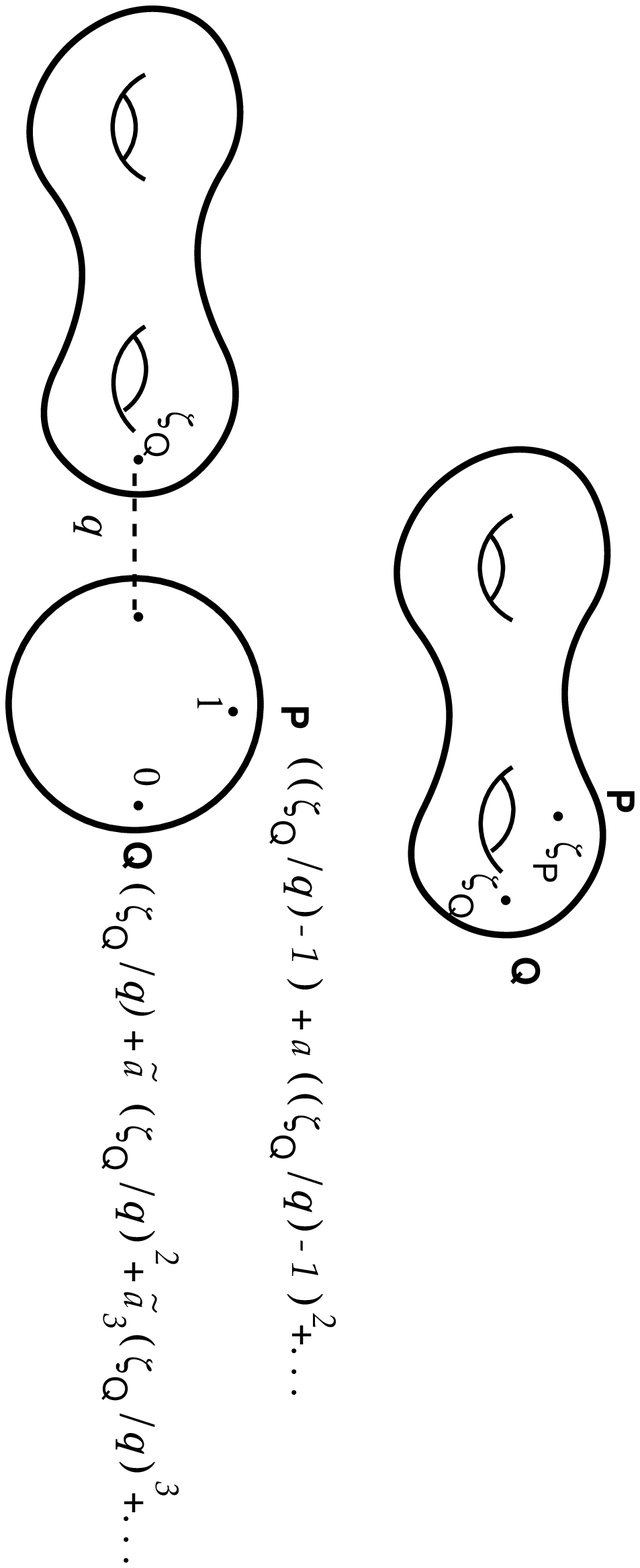}{90}{55}{55}{425}{-130}

Finally, the above formalism can be extended to the heterotic string.
There we are interested in integrating over the moduli space of
super-punctured Super-Riemann surfaces, $\mh gs$.
Again, a measure on $\mh gs$ can be constructed by introducing the
augmented moduli space $\ph gs$ of punctured surfaces with superconformal
coordinates at the punctures and using a section to pullback the
naturally defined measure.
The heterotic string contains new features that do not
appear in either the semi-rigid or bosonic cases and so we will develop
the necessary formalism in more detail later in the paper.

\newsec{Bosonic Dilaton Two-Point Functions}
In the bosonic string, we are able to
come up with a `good' coordinate slice for all of the moduli space
$\mgs 04$ and
use this to calculate the two-point functions on the sphere of dilatons
and strong physical states. The nice feature is that we can give a
global family and not just restrict ourselves to a calculation in one
patch, as was done in \TCCT.
This will also give us a basic building block from which coordinate
families appropriate for higher genus surfaces can be constructed.
\subsec{The Geometry for the Sphere}

We are interested in computing two-point functions on the sphere. To
this end, we choose a convenient coordinate slice for the four-punctured
sphere as follows: The sphere needs two coordinate
patches to cover it. The northern hemisphere has coordinate $z$, while
the southern hemisphere has coordinate $w=-1/z$.
 Now, letting $\rr$ (resp. $r$) be the modulus for the location
of the point $P$ (resp. $Q$), we can linearly
interpolate between the two hemispheres across the equator by letting
$h(|\rr |)$ (resp. $g(|r|)$) be any function that smoothly interpolates
between 1 and 0 as $\rr $ (resp. $r$) goes from 0 to $\infty$,
\tfig\interpol.
Then
\eqn\zetaeq{\eqalign{\zeta_P (\cdot)&= h (z-\rr) +
                              (1-h) \rr ^2 (-1/z +1/\rr)\cr
 \zeta_Q (\cdot) &= g (z-r) +(1-g) r^2 (-1/z +1/r)}}
give coordinates centered at $P$ and $Q$ respectively, and appropriate for
both hemispheres. The curvature of the sphere manifests itself through
the non-holomorphic behavior of the interpolating functions $g$ and $h$.
This is fine as
long as $|r-\rr|> \epsilon$ for some $\epsilon >0$. But in the
neighborhood of $P$ and $Q$ close together the coordinates should go over
to ones that represent the conformally equivalent picture of a sphere
containing the two punctures pinching off from the rest of the surface.
\insfig\interpol{The interpolating functions $f$ and $g$. $h$ has the
same behavior as $g$, but its argument is $\rr$. Any smooth functions
that run between 0 and 1 could be used.}{2.5}
{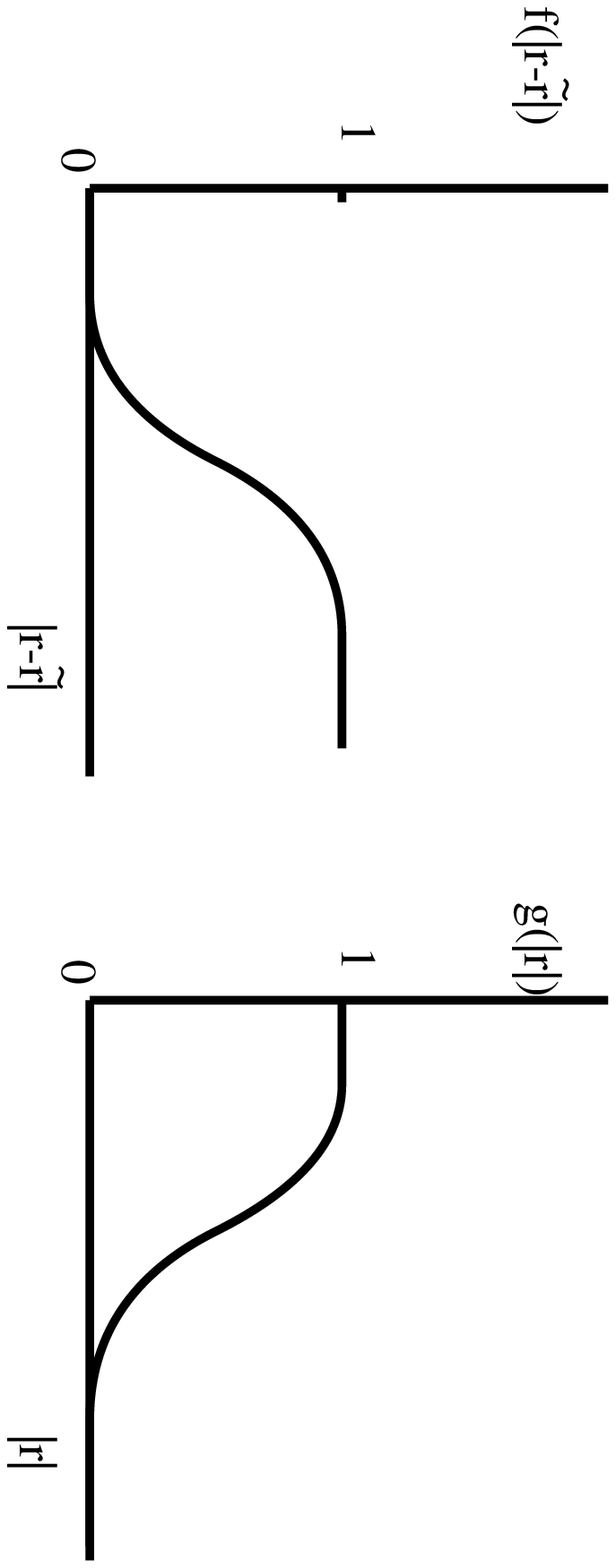}{90}{55}{55}{420}{-110}

The plumbing fixture is the standard construction to represent the
pinching off of a sphere from the rest of a surface. Computing contact
terms requires a three-punctured sphere with coordinates pinching off
from the rest of the sphere. In \TCCT a physically motivated
choice for choosing appropriate coordinates was given.  We will use the
more general choice given in \DESRST with
the three punctures placed at 0 where the coordinates are
\eqn\zercor{\xi +\aa \xi^2+\aa_3\xi^3 +\cdots,} at 1 with coordinates
\eqn\onecor{(\xi-1) +a (\xi-1)^2+ a_3(\xi-1)^3 +\cdots,}
and at $\infty$ with coordinate $1/\xi$.
We choose the most general coordinates that vanish at the
points $P$ and $Q$ since this is where our operators will be inserted.
The coefficients $a$ and $\aa$ in the expansions of the coordinates will
be determined by requiring that dilaton-strong physical state contact
terms have the correct normalization; the higher coefficients drop out in
dilaton calculations.
The point at $\infty$ is sewn via the standard plumbing fixture onto the
point Q by the identification
\eqn\bsew{\xi = \zeta_Q (\cdot)/q. }
So for small $|r-\rr|$ the coordinates at $P$ and $Q$ in \onecor
and \zercor become
\eqn\phieq{\eqalign{\phi_P (\cdot) &= (\zeta_Q (\cdot)/q-1)+
                    a (\zeta_Q (\cdot)/q -1)^2 + \cdots \cr
    \phi_Q (\cdot) &= \zeta_Q (\cdot)/q+\aa (\zeta_Q (\cdot)/q)^2 +
                                  \cdots\,.}}
Essentially the pinching parameter $q$ and $r$ have replaced $\rr$ and
$r$ as the moduli describing the locations of the two points. The
relationship between $q$ and $\rr$ and $r$ can be found by
demanding that $\phi_P(P)=0$. Thus,
\eqn\qeq{q= \zeta_Q(P)= {{(\rr-r) ((1-g) r +g \rr)}\over{\rr}}. }

Finally, the two coordinate systems are joined by
using yet another function $f(|r-\rr|)$ that smoothly interpolates from 0
to 1 as $|r-\rr |$ goes from 0 to $\infty$, \interpol.  Again we choose
a linear interpolation. It is important to realize that we cannot interpolate
between just any two coordinates because there is a phase that cannot be
removed. (Hence the factors of $r^2$ and $\rr^2$ in \zetaeq .) The phase is
found by expressing $\zeta_Q$ in terms of $\zeta_P$ and rewriting
$\phi_P$ in terms of $\zeta_P$. The final result is
\eqn\sigeq{\eqalign{
   \varphi&={{((g-1) r^2 -g \rr^2)} \over {\rr(r-\rr)((1-g) r+g \rr)}}\cr
\sigma_P(\cdot) &= f \varphi \zeta_P + (1-f) \bigl(\varphi \zeta_P
           +(a \varphi^2 - {{h (1-g) r^2}\over{q \rr^3}}+
             {{g (1-h)}\over{q \rr}}) \zeta_P^2+\cdots\bigr)\cr
\sigma_Q(\cdot)&= f(\zeta_Q/q) +(1-f)\bigl(\zeta_Q/q + \aa (\zeta_Q/q)^2
                     +\cdots\,\bigr)}}
where $\varphi$ is the relative phase for the coordinates at $P$.
These are the coordinates for the holomorphic sector. The
antiholomorphic coordinates are simply the barred version of this.

As a warm-up (and because we will need it later) we will calculate the
one-pont function of the dilaton on the sphere using the linear
interpolation in \zetaeq. A different interpolation was used in \TCCT.
Since we are only
inserting one operator, the $\zeta_Q(\cdot)$ coordinate alone is all that is
needed. The pushforwards are calculated by differentiating the coordinate
with respect to $r$ and $\rb$ and using
\eqn\pushform{\eqalign{
  \zeta_{Q*}({\dbd r})&={\der {\zeta_Q}{r}} {\dbd {\zeta_Q}} +
                  {\der {\bar \zeta_Q}{r}} {\dbd{\bar \zeta_Q}}\cr
   \zeta_{Q*}({\dbd \rb})&={\der {\zeta_Q}{\rb}} {\dbd {\zeta_Q}} +
                  {\der {\bar \zeta_Q}{\rb}} {\dbd{\bar \zeta_Q}}.}}
Taking the derivatives and re-expressing the result in terms of
$\zeta_Q$ gives
$$\eqalign{\der {\zeta_Q}{r}&=-1 +{2\over r} (1-g)\zeta_Q -
      {2 g (1-g)\over r^2} \zeta^2_Q +{|r| \gp \over 2 r^2} \zeta^2_Q +
       \cdots \cr
     \der {\zeta_Q}{\rb}&={\gp \over 2 |r|}\zeta^2_Q+\cdots\, .}$$
Thus the corresponding $b$-insertions are
$$\eqalign{b[\zeta_{Q*}({\dbd r})]&=b_{-1} +\cdots\cr
           b[\zeta_{Q*}({\dbd \rb})]&=-{\gp\over 2 |r|}b_{1} +\cdots}$$
and, hence, the one-point function is given by
\eqn\onept{\eqalign{\vev{D}=&
\int {\dd r\we\dd\rb\> \bra{\Sigma}b_{-1} \left(-{\gp\over 2 |r|}
              \right)b_{1} (-2) c_{-1} c_{1}\ket{0}}\cr
=& \int{\dd r\we\dd\rb\> { \gp \over|r|}} Z\cr
=& -2 i\int{\dd |r|\we\dd\th \>\gp} Z\cr
=& -2 \pi i\,(-2) Z\; ,}}
where we recall that $g$ runs from 1 to 0 as $|r|$ runs from 0 to
$\infty$.
($Z$ is the partition function on the sphere and the insertions
for the three conformal Killing vectors have been suppressed.)
This is just what was expected since the sphere has $2g-2=-2$.
Also note that, as in \TCCT,  the form of $g$
didn't matter and that the measure was a total derivative.
Of particular importance for our later calculations is
the fact that the support of the measure is compact. For it is restricted to
the region in which $g'$ is non-zero. This region can be chosen to be an
annulus in the $r$-plane, and the measure is
identically zero away from this annulus. Thus when integrating, we can
use Stokes' Theorem and contributions will come from the boundary of
the annulus. Similar things will happen in the two-point function
calculations.

\subsec{Two-Point Functions on the Sphere}
The computation for the two-point functions is more or less identical to
the one for the one-point function. However, now there are four
pushforwards to compute and we must use the $\s_P$ and $\s_Q$
coordinates in~\sigeq.
The resulting $b$-insertions are \foot{The derivatives and algebra are
straightforward but lengthy. These pushforwards (and the heterotic ones
also) were calculated with the help of {\it Mathematica.}}

\eqn\bins{\eqalign{\bp {r}=&{1\over q} \b{-1}{Q}+ \cdots\cr
  \bp {\rb}=&\biggl({a \y \over 2 \ay} \fp+{h m \y^2 \over 2 n \ay} \fp-
   {g m\rr^2 \y^2 \over 2 n^2 \ay} \fp +\cr &{(1-f) r^3 \rr^2 \y m \over
    2 n^3 |r|}\gp+{a (1-f) r^2 \rr \y \over 2 m n |r|}\gp\biggr)
 \b{1}{P}+ \cr
 &\biggl({\aa \y \over 2 \ay} \fp - {\aa (1-f) r \y \over 2 m |r|} \gp+
    {m \y \over 2 \rr |r|} \gp\biggr) \b{1}{Q} +\cr
& {1 \over {\bar q}} \bb{-1}{Q}+ \cdots\cr
  \bp {\rr} =& \varphi \b{-1}{P} +\cdots \cr
\bp {\rrb}=& \biggl(-{a \y \over 2 \ay } \fp -
{h m \y^2 \over 2 n \ay } \fp +
      {g m\rr^2 \y^2 \over 2 n^2 \ay} \fp -\cr &
       {f m \rr \y \over 2 n |\rr|} \hp \biggr) \b{1}{P}+
       \bar \varphi \bb{-1}{P} -
      {\aa \y \over 2 \ay} \fp \b{1}{Q}+\cdots}}
where we have defined
\eqn\mndef{\eqalign{m=&(1-g) r +g \rr \cr
                    n=&(1-g) r^2 +g \rr^2 \cr}}
for convenience (and $\bar \varphi$, resp. $\bar q$, is the complex conjugate
of $\varphi$ , resp. $q$). The derivatives are all with respect to the argument
shown. And, finally, the dots represent $b$-insertions that do not
contribute to the dilaton-SPS and dilaton-dilaton two-point functions.

We will calculate the dilaton-SPS in two ways. The first is by putting
the dilaton at $P$ and the SPS at $Q$, and the second by switching them.
With the dilaton at $P$ and the SPS at $Q$, we need to pick off the
contribution of the measure proportional to
$\b{-1}{P} \b{1}{P} \b{-1}{Q} \bb{-1}{Q}$:
\eqn\dsmeas{\eqalign{\bp{\rr}\bp{\rrb}\bp{r}\bp{\rb}=&\cr
   {1\over |q|^2}\bigl({a n \fp\over 2 \rr m \ay} -
              {g \rr \y \fp\over 2 n \ay} -
   &{f\hp\over 2 |\rr|}
 +{h \y\fp\over 2 \rr \ay}\bigr)\cr
      &\times\b{-1}{P} \b{1}{P} \b{-1}{Q} \bb{-1}{Q}}
}
and, hence, we need to integrate
\eqn\dsfouri{\eqalign{
\int&\vf{1\over|q|^2}\biggl({a n \fp\over 2 \rr m \ay} -
     {g \rr \y \fp\over 2 n \ay} - {f\hp\over 2 |\rr|} +
\cr&{h \y\fp\over 2 \rr \ay} \biggr)
              \biggl({1\over\varphi}\biggr)^{L_0^P}
              \biggl({1\over\bar{\varphi}}\biggr)^{\bar{L}_0^P}
       q^{L_0^Q} \qb^{\bar{L}_0^Q}
 \b{-1}{P} \b{1}{P} \b{-1}{Q} \bb{-1}{Q}\ket{D}^P\otimes\ket{\psi}^Q
}}
over the moduli space. The factors of $(1/\varphi)^{L_0}$ and $q^{L_0}$
are needed because we
have inserted our states using \sigeq, and we would like to compare this
with the insertions of states in the coordinates $\zeta_P$ and $\zeta_Q$
in \zetaeq \TCCT. The dilaton
has $L_0=\bar{L}_0=0$ but the state $b_{-1} \bar{b}_{-1}\ket{\psi}$ has
$L_0=\bar{L}_0=1$, leaving
\eqn\dsfour{\eqalign{
\int\vf\biggl(&{a n \fp\over 2 \rr m \ay} -
     {g \rr \y \fp\over 2 n \ay} -
   {f\hp\over 2 |\rr|} +\cr&{h \y\fp\over 2 \rr \ay} \biggr)
 \b{-1}{P} \b{1}{P} \b{-1}{Q} \bb{-1}{Q}\ket{D}^P\otimes\ket{\psi}^Q
}}
The above can be written
$$\int\vf{\dbd{\rrb}}\biggl(-{a n f \over \rr m \y} -{h f\over \rr} +
                {g \rr f\over n}\biggr)
 \b{-1}{P} \b{1}{P} \b{-1}{Q} \bb{-1}{Q}2 c_1 c_{-1}
\ket{0}^P\otimes\ket{\psi}^Q, $$
or, erasing the $\ket{0}$ at P,
\eqn\dtf{\int\dd \biggl[2\biggl({a n f \over \rr m \y} +{h f\over \rr} -
                {g \rr f\over n}\biggr) \tfa\biggr]
\b{-1}{Q}\bb{-1}{Q} \ket{\psi}^Q.}
With care this form can be integrated using Stokes' Theorem.
The purpose of the interpolation in \sigeq is to smooth out the
delta-function contact term at $\rr=r$. In fact, the function
$f(|r-\rr|)$ acts as a small distance cut-off and we should
change to the more natural variable $y=\rr-r$ in place of $\rr$.
Then we can integrate over $y$ leaving
us with a two-form corresponding to the insertion of the SPS.
Making the change of variables in \dtf gives
\eqn\yform{\eqalign{\int\dd \biggl[
       2\biggl(-{a \nsub f(|y|) \over (y+r) y \msub}&+
 {h(|y+r|) f(|y|) \over y+r} -\cr {g (y+r) f(|y|) \over \nsub}&\biggr)
     \yfa\biggr] \b{-1}{Q}\bb{-1}{Q} \ket{\psi}^Q.}}
To apply Stokes' Theorem it is necessary to determine the support of the
original four-form in \dsfour and what the appropriate boundaries are for
each term in the three-form. As in the one-point calculation, the form
we have to calculate is compactly supported on several annuli.
Each term in \dsfour is either proportional
to $\fp$ or $\hp=\hpy$. The support of $\fp$ is confined to the
small annulus around $y=0$ shown in \tfig\conti. Thus terms in
\yform that come from terms in \dsfour proportional to $\fp$ should be
integrated around the boundaries of this annulus and only a pole at
$y=0$ can contribute. Furthermore, it should be noted that $f=0$ on the
inner boundary and $f=1$ on the outer.
The support of $\hp$ is centered around $\rr=0$ and so in the $y$-plane
this is an annulus around the point $y=-r$, also depicted in \conti.
Since $h=1$ for small $\rr$ and $h=0$ for large $\rr$, the inner contour
has $h=1$ and the outer $h=0$.

\insfig\conti{Regions of integration in the $y$-plane. The forms that
are being integrated only have support in the annular regions. Stokes'
Theorem is used to rewrite them as contour integrals.}{2.5}
{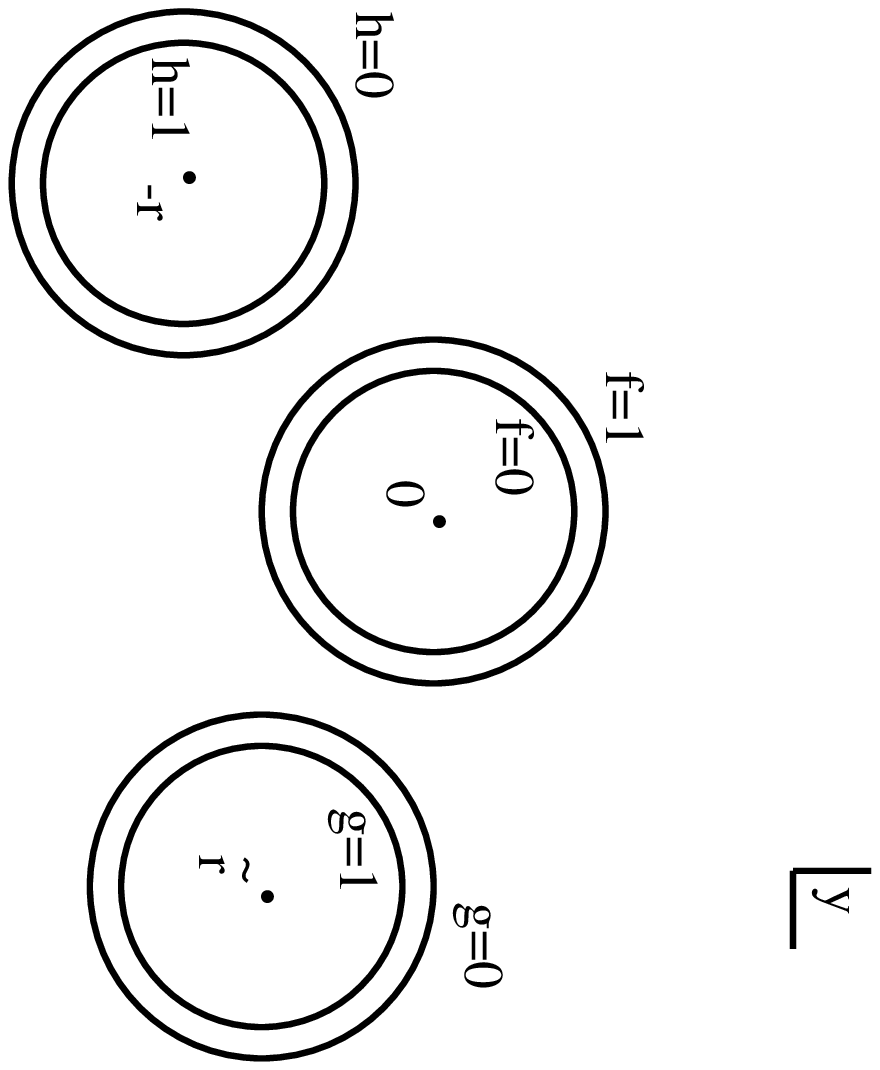}{90}{55}{55}{420}{-130}

We can now apply Stokes' Theorem. The first and third terms
\yform come from something
proportional to $f'(|y|)$ in \dsfour and, hence, have support only on
the small
annulus around the origin in the $y$-plane.
The middle term comes from
something proportional to both $f'(|y|)$ and $h'(|y+r|)$ and so has
support on both the annulus around the origin and the
annulus around the point $y=-r$. Hence, poles at 0 and $-r$
contribute for this term. The integrals are now straightforward and
the result is
\eqn\dspsi{-2 \pi i(-2 a -2 )\int\tf\b{-1}{Q}\bb{-1}{Q}\ket{\psi}^Q.}
This corresponds to
\eqn\dsps{\vev{D \psi}=-2 \pi i (-2 a -2 )\vev{\psi},}
which is the dilaton equation on the sphere if we choose $a=-1/2$. This
is the residual choice of coordinates mentioned earlier.

Since our plumbing fixture was asymmetric, we should see what happens if
we put the dilaton at $Q$ and the SPS at $P$. This isn't really an
independent check, since we will determine $\aa$ instead of $a$. The
real question is will the resulting values of $a$ and $\aa$ will give
the right contribution for two dilatons on the sphere.

The piece of the measure proportional to
$\b{-1}{P}\bb{-1}{Q}\b{-1}{Q}\b{1}{Q}$ is
$$|\varphi|^2
  \biggl(-{\gp\over 2 |r|}-{\aa \rr \fp\over 2 m \ay}+
            {\aa (1-f) r \rr \gp \over 2 m^2 |r|}\biggr)
\b{-1}{P}\bb{-1}{Q}\b{-1}{Q}\b{1}{Q} $$
which again gives rise to a total derivative (after changing variables
to $y$ in favor of $r$ ($\rr$) in the first (second) term), with the
integral now
$$\eqalign{
\int\biggl(\dd \biggl[2\biggl({\aa g (1-f) \rr\over (\rr+y (g-1) (\rr-y)}
            +{g \over (y-\rr)}\biggr)&\yfb \biggr]\quad+\cr
          \dd \biggl[2\biggl(-{\aa (1-g) f (y+r)\over \msub r}
                   +{\aa f (y+r)^2\over \msub y r}\biggr)&\yfa\biggr]
\biggr)\b{-1}{P}\bb{-1}{P}\ket{\psi}^P.}$$
This time the factor $|\varphi|$ was cancelled by the factors of
$(1/\varphi)^{L_0}$ and $(1/\bar{\varphi})^{\bar{L}_0}$ since the
strong physical state is located at $P$. The integrals are again
straightforward once it is noted that $\gpy$ has support on an annulus
centered at $y=\rr$, as shown in \conti. It should be noted that in the
first term integrating over $y$ leaves a two-form
proportional to $\dd \rr\we\dd\rrb$ corresponding to the insertion of
the SPS at the point $P$, while integrating the second term leaves us
with a two-form proportional to $\dd r\we\dd\rb$. However, this is not
a problem since the second term is purely a contact term and appears
only when the two points coincide. With this in mind, we find
\eqn\spsd{\vev{\psi D}=-2 \pi i (2 \aa -2)\vev{\psi}}
and we have to choose $\aa=1/2$ to get the dilaton equation. These
choices of $a$ and $\aa$ agree with the physically motivated ones in
\TCCT, although one must be careful about signs when comparing the two
calculations.

The two-dilaton calculation is similar to the above calculation, only
more involved. Because of its length, we have moved it to Appendix A and
only give the result here. One can readily recognize the
structure of each term  in the two-point function. We see
again that we have a total derivative and that, after integrating
over $y$, we are left with the measure for the insertion of one dilaton
which we recognize from the one-point calculation. The resulting form
can be completely, unambiguously integrated. The result is
\eqn\dildil{\eqalign{\vev{DD}&=-2 \pi i (2(\aa-a+a \aa)-2)\vev{D}\cr
                            &=-2 \pi i (3/2-2)\vev{D}.}}
The $3/2$ spoils the dilaton equation and agrees with the
calculations in \TCCT and \PRIV, which were done along the lines that we
will use for the heterotic string.

As mentioned earlier, this failure of the dilaton equation is somewhat
mysterious. Comparing \dildil and \dsps, we see
that the 3/2 results from extra terms proportional to $\aa$, or terms
that depend on the coordinates at $Q$ on the three-punctured sphere. In
the heterotic case, terms like this are present, but they are killed by
the integration over the odd moduli. This is also the case with
potentially divergent contributions that are akin to the tachyon
divergence. The
low-energy theorem for the zero-momentum dilaton usually identifies it
with string coupling constant. More explicitly, inserting a dilaton
is supposed to correspond to $\lambda \dbd{\lambda}$. We have seen that
$$\vev{DO_{i_1} \cdots O_{i_n}}=(2 g-2 +n)\vev{O_{i_1} \cdots
O_{i_n}},$$ where the $O_i$ are SPS's. Thus, as expected, the insertion
of one dilaton corresponds to the first variation with respect to the
string coupling constant. However, we have also seen that when two
dilatons are inserted the result is
$$\vev{DDO_{i_1} \cdots O_{i_n}}=(2 g-2 +n+3/2)\vev{DO_{i_1} \cdots
O_{i_n}}.$$ If the insertion of a second dilaton were to correspond to
the second variation of the string coupling constant, we would have
expected a 1 in place of the 3/2. Possibly, this surprising result
is in some way due to the tachyon. As further support for this, we will
find that the heterotic string behaves nicely, with no tachyon
divergence and with a 1 instead of a 3/2.

\newsec{`Good' Coordinate Families on Higher Genus Surfaces}

In the previous section, we were able to calculate two-point functions
on the sphere because we were able to provide a `good' coordinate family
for the moduli space $\mgs 04$. This section is devoted to sketching how
this can be used as a building block to provide a `good' coordinate
family on a higher genus surface.
\foot{I would like to thank Jacques Distler for pointing this out.}
We begin by examining the what made the
coordinate family given in \sigeq `good'. For large $q$ (and large
$|r-\rr |$) the points $P$ and $Q$ are widely separated and the coordinate
$\sigma_P(\cdot)$ becomes independent of the moduli for $Q$ (except for
the $r$ dependence in the overall phase which is irrelevant here).
In this region, the coordinates look like coordinates one would choose
if $P$ were the only puncture on the sphere. These coordinates
interpolate between the coordinates $z$ which is good for everywhere but
the south pole and $-1/z$ at the south pole. On the other hand,
$\sigma_P(\cdot)$ is constructed to go over to the plumbing fixture
coordinates when the $P$ and $Q$ approach each other. In this region,
the coordinate for $P$ depends essentially on the moduli associated
to the location of $Q$. Finally, since
the sphere has no moduli, $\sigma_P(\cdot)$ is trivially independent of
the moduli associated to the unpunctured surface.

We would
like to now generalize this coordinate family to higher genus surfaces.
The key is to decompose the surface via a `pants' decomposition into a
set of three-punctured spheres on which we can easily give `good'
coordinate families for the insertion of another puncture.
Then, we must show that we can glue these local coordinate families
together into a global one that covers all of moduli space. Thus we
will demonstrate the existence of a `good' coordinate family that
was presumed to exist in the one patch calculations in \TCCT and \DESRST.
One could then go on to calculate dilaton correlation functions on
higher genus surfaces.

A `pants' decomposition of an unpunctured genus $g$ surface
is accomplished by
choosing a maximal set of $3g-3$ non-intersecting closed geodesics on
the surface, as shown in \tfig\pants.
These curves decompose the surface
into $2g-2$ pants-shaped regions which can be thought of as
three-punctured spheres. The original surface can be reconstructed by
sewing together these spheres using the plumbing fixture construction.
In this section, it helpful to use a modified plumbing fixture. Instead
of joining together $z_1$ and $z_2$ by $z_1 z_2=q$ as we did earlier, we
will take $q$ to be a modulus of a two-punctured sphere and, choosing
the coordinate near the north pole of the two-punctured sphere $w$ and
the coordinate near the south pole $q/w$, sew it between the two
three-punctured spheres by $z_1 w =1$ and $z_2 q/w=1$. Thus when a
puncture $P$ on one of pants-shaped regions approaches a boundary,
it moves from a three-punctured sphere onto the two-punctured sphere
that sews it to its neighbor. Thus an isolated region of the surface can
be represented as in \tfig\tsph.
\insfig\pants{Pants decomposition of a genus three surface.}{1.5}
{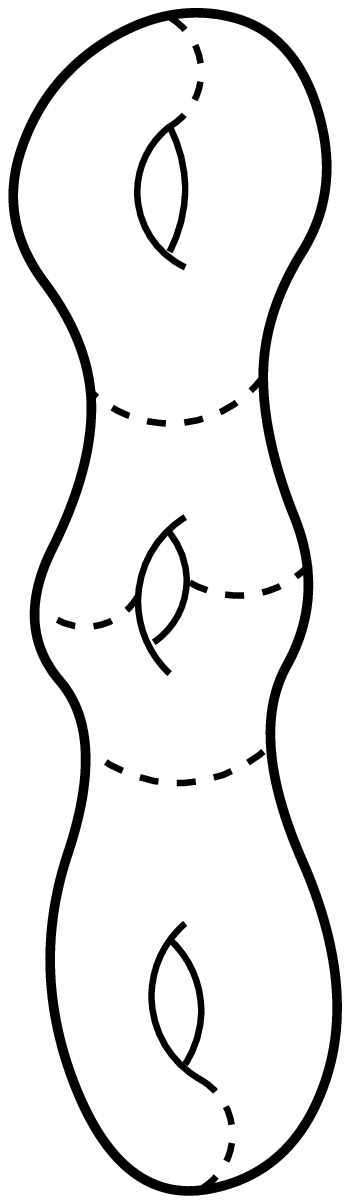}{90}{55}{55}{420}{-130}
\insfig\tsph{Sewn spheres in one region of a pants-decomposition. The
three-punctured sphere is one of the pants-shaped regions of the
surface, while the two-punctured spheres contain the moduli of the
plumbing fixtures that sew the region to the rest of the surface. The
`x' represents an insertion. As the location of the insertion moves into
one of the dotted circles, it moves through the plumbing fixture onto
the adjacent sphere.}{4.0}
{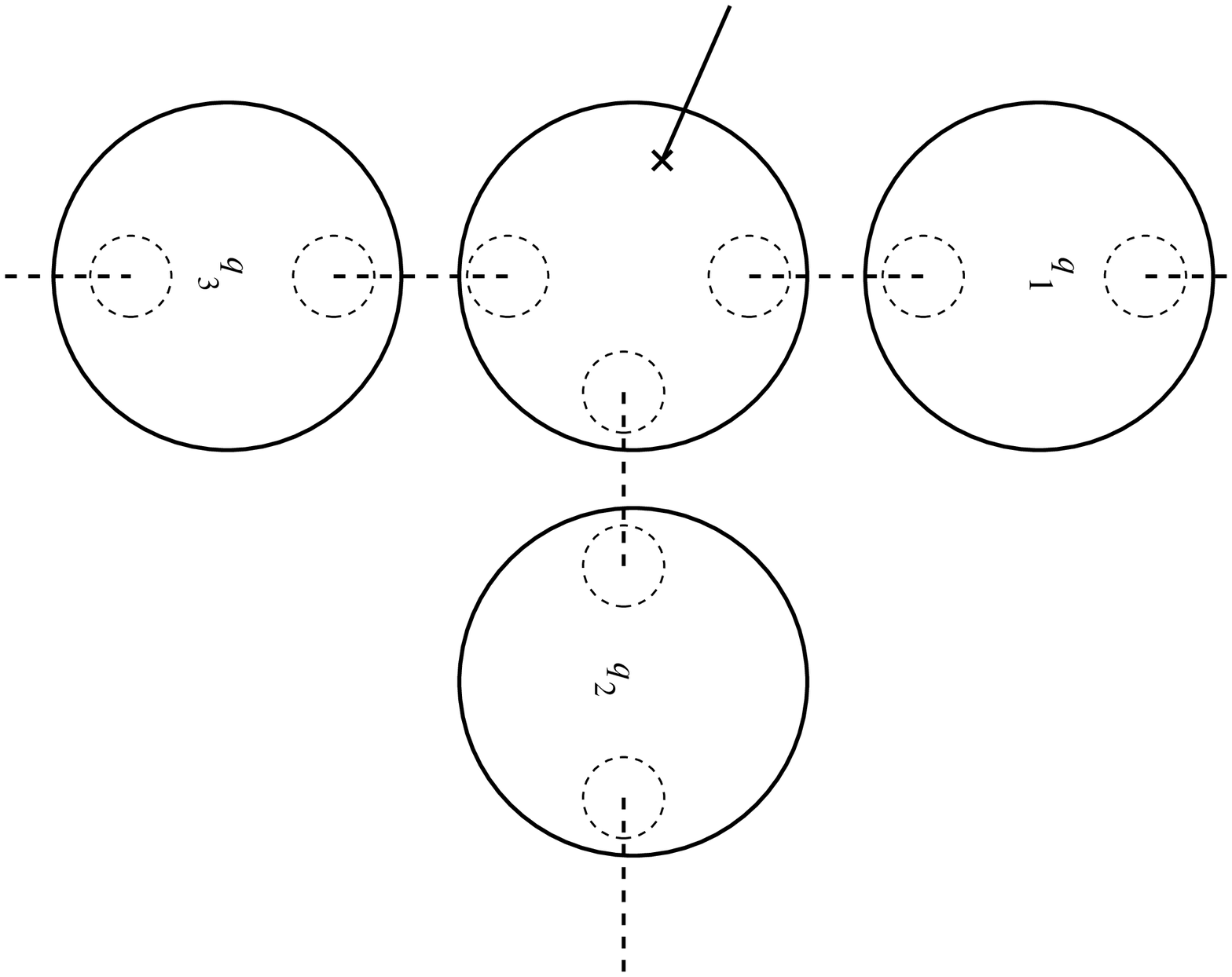}{90}{35}{35}{350}{10}
Handles are created by sewing together two punctures on the
same sphere. Varying the $q_i$ in the plumbing fixtures then corresponds
to varying the $3g-3$ moduli of the surface. Actually, we have to
consider all distinct pants decompositions of a surface, each giving
rise to a cell in moduli space. The boundaries between the cells can be
thought of as corresponding to replacing two three-punctured spheres by
one four-punctured sphere. This is depicted for a genus two surface in
\tfig\genustwo.
\insfig\genustwo{Different cells of a pants decomposition for a genus two
surface. The dotted lines in the upper diagrams represent the
sewings that correspond to the pants-decompositions shown in the
lower diagrams. (The two-punctured spheres for the sewing have been
suppressed.) The middle diagram is the boundary between the two
decompositions.}{2.8}
{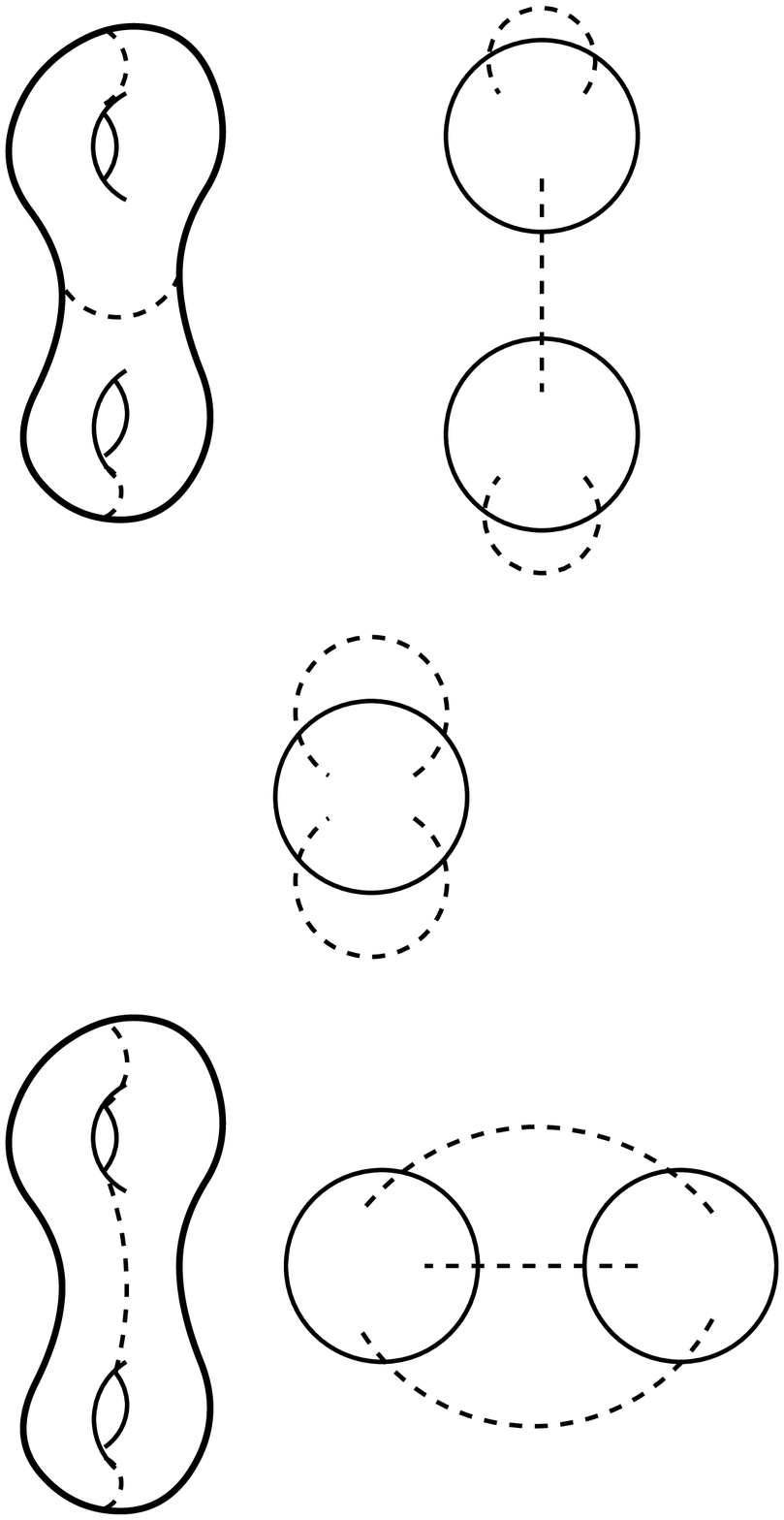}{90}{50}{50}{420}{-20}

With these preliminaries out of the way, we can now start to think
about integrating a single puncture over a surface. To do this, we have
seen that we need a `good' coordinate family over the entire surface. The
pants-decomposition has provided us with a set of punctured spheres all
sewn together. We will thus be able to integrate
the puncture over the entire surface if we can provide a `good'
coordinate family over each of the punctured spheres that behaves nicely
as we move from one sphere to another, and if the
resulting family is continuous across the cell boundaries of moduli
space (different pants-decompositions correspond to different cells).
The first part can be accomplished by an easy generalization of the
coordinates used in section 3. And the second part will require a `good'
coordinate family on $\mgs 0n$. But this too is an easy generalization
of the coordinate family on $\mgs 04$.

For simplicity, we will focus on a sphere on which
each of its punctures is sewn to a puncture on another sphere. Our
discussion could be expanded to include the case where two punctures
on a sphere are sewn together to form a handle. Our sphere is sewn to
three two-punctured spheres each of which contain one of the moduli of
the surface (i.e., a $q_i$). When the puncture $P$ is far from one of
the sewn punctures, its coordinates are the same as before:
\eqn\threepunc{
 h(|r|)(z-r)+(1-h(|r|)) r^2 \bigl({-1\over z}+{1\over r}\bigr).}
This is independent of the moduli of the surface.
On the other hand, when $P$ approaches any one of the sewn
punctures, we can think of it as having moved through the neck of the
plumbing
fixture onto a two-punctured sphere. In this region the coordinate is
\eqn\twopunc{
 g_i(|r_i|)(w_i-r_i)+(1-g_i(|r_i|)) r_i^2
  \bigl({-q_i\over w_i} +{1\over r_i}\bigr),}
and it depends on one of the $q_i$. It should be noted, however, that
even in this region the coordinates are independent of most of the
surface's moduli. In \twopunc,
$w_i$ is a coordinate for one of the $3g-3$ two-punctured spheres, and
$r_i$ is the location of the puncture on this plumbing fixture
sphere. Finally, to get a `good' coordinate family in this region, we
just have to interpolate between the all of the different behaviors. So
there will be three interpolating functions that will depend on the
differences between $r$ and positions of the three sewn functions. Then
the coordinate family can be patched together using these functions;
each interpolation will connect a family like the one in \threepunc to
one as in \twopunc.
This is somewhat messy to write down, but the underlying idea should be
clear. One should picture the puncture wandering around on a
three-punctured sphere and moving over to one of the two-punctured
spheres through the neck of a plumbing fixture when it wanders too close
to one of the sewn punctures.

Another essential ingredient is a coordinate family appropriate for
$\mgs 0n$.
This is needed for both when there are other operators inserted
onto the surface and for demonstrating that our family is continuous
across the cell boundaries in moduli space. But this presents no new
problems. We proceed as before, except that now we have to interpolate
whenever any one of the $n$ punctures approaches another. Again this
would be messy to write down, but the principle is clear enough.
The important point is that a `good'
coordinate family can be written done as before, one in which the
coordinates for our inserted puncture is independent of the moduli for
the other punctures when it is far from them, and in which, as two
punctures collide, it goes over to the coordinates given by the plumbing
fixture.

As mentioned earlier, the cell boundaries can be represented as a
transition between two different pants-decompositions through a
multi-punctured sphere. Our coordinate family behaves smoothly through
this transition and it always remains `good', so we have a prescription
that works globally on the moduli space. Finally, if there were other
punctures on the surface, then we could just use the coordinate family
for $\mgs 0n$ on the many-punctured spheres that now result from the
pants-decomposition. This would have extra interpolations as the
puncture $P$ neared the other punctures.

To recap, our prescription for constructing a `good' coordinate
family on a surface for inserting a puncture is given by  first
decomposing the surface into a set of punctured spheres that are
sewn together by using a plumbing fixture with two-punctured spheres.
Coordinates are then chosen by merely interpolating between the
coordinates appropriate for the different regions of the surface
as the point $P$ wanders around the surface and appropriate for
regions where $P$ approaches other punctures. By construction, the
coordinate family is independent of the moduli that are in some sense
`far' from $P$. The family is continuous between cell boundaries because
it is well-behaved on the many-punctured spheres that provide the
transition between different pants-decompositions.

One could now go on and calculate dilaton correlation functions. It is
clear that they would go through as for the two-point function on the
sphere. Once again the measure will be compactly supported in the
regions where the interpolations are taking place and Stokes' Theorem
will be easily applied. With this sketch of how one could put the
bosonic calculations on firmer footing complete, we now turn to
dilaton contact terms in the heterotic string.

\newsec{Dilaton Contact Terms in the Heterotic String}

\subsec{The Geometry}
Now let us turn to the heterotic case. Once again we adopt the viewpoint
of the contact interaction as being the degeneration of a surface with
the points corresponding to our two operators pinching off from the
rest of the surface. The coordinates for this region of moduli space are
now provided by the standard heterotic plumbing fixture. A new feature
is that the three-punctured super-sphere is no longer rigid: it has one
odd modulus associated to it. Again we will match the coordinates
appropriate for this region onto the coordinates appropriate away from
the compactification divisor. In this case, we will match up to a
superconformal normal-ordered family of coordinates. We will then calculate
the pushforwards necessary for the construction of a good string measure.

Before proceeding, we will give a brief review of the necessary
geometry. For a more complete treatment see \LSMS. We recall that a
super-Riemann surface (SRS's) can be constructed via a patch definition.
That is, locally our surface looks like a region of the
${\bf C}^{1\vert 1}$ plane for which we use the coordinates
$(z, \theta ; \bar z , \bar \theta)$. Since we are interested in the
 heterotic string we will
only consider the holomorphic sector and we can always obtain the
antiholomorphic sector by taking the complex conjugate and setting all
odd variables to 0. These local patches can now be glued together to
build up a SRS using superconformal transition functions. These are
transformations under which the super-derivative $D_{\theta} =
\pa _{\theta} + \theta \pa _z$ transforms homogeneously. This requirement
means that the new $z', \theta '$ must satisfy
\eqn\sct{  D_\theta z'=\theta ' D_\theta \theta' . }

To calculate correlation functions it is necessary to introduce
punctures on the SRS. In this paper we need only consider super-punctures
and not spin punctures since we will only consider operators in the
Neveu-Schwarz sector.
In fact, in order to construct a measure, we will
need our punctures to come equipped with a superconformal coordinate
that vanishes at the point. We can thus specify a point on the SRS by
giving an even function $\s =\s (z, \th )$ and recover the puncture as the
point where $\s $ and $D_\th \s $ vanish. Given a coordinate $\s$ we can
also construct its odd partner, which we denote by $\sth $, by requiring
that $\s$ and $\sth$ are superconformally related to $z$ and $\th$, i.e.,
$$  D_\th \s = \sth D_\th \sth $$ holds. Then the puncture is given by
the vanishing of $\s$ and $\sth\,$
\foot{The vanishing of $\sth$ is equivalent to the vanishing of
$D_\th \s$ for they are related by the multiplication of a non-vanishing
function. However, it is important that we use $\sth$ and not $D_\th \s$
as the coordinates of the point in order to maintain superconformal
invariance.}.
For completeness, we note that if
\eqn\sci{\s(z,\th) =f(z)+\th \alpha (z),}
then
\eqn\scii{\sth(z,\th) =\beta (z)+\th g(z)}
with
\eqn\sciii{\eqalign{g(z)^2=&f'(z)\cr
 \beta (z)=&{\alpha (z) \over g(z)}.}}
(This is true as long $\alpha(z)$ and $\alpha'(z)$ are proportional to the
same odd parameter, which will always be the case for us.)
The prime denotes differentiation with respect to $z$. We will sometimes
denote the location of the puncture as $[\s]$ and we will always just
give $\s$ as the coordinates of the point, it being understood that we mean
the pair $(\s,\sth)$. If we choose $\s=(z-u+\xi \th)$,
then we call such a family `superconformal normal ordering' (SCNO)
\EFEFS.

The moduli space of a genus $g$ SRS with $s$ super-punctures is denoted by
$\mh gs$ and is $(3 g-3 +s \vert 2g-2 +s)$ dimensional. We also need the
augmented moduli space, $\ph gs$, which is the infinite dimensional
moduli space of punctured
SRS's with a choice of superconformal coordinate centered at each
puncture. There is a natural projection $\pi : \ph gs \to \mh gs$ given
by forgetting the coordinate at the puncture. A coordinate family is
then given by a slice $\s : \mh gs \to \ph gs$.
The measure on $\mh gs$ is obtained by using $\s$ to pull back a measure
that is naturally defined on $\ph gs$. The measure is defined in a way
similar to \meas for the bosonic case, except here we have odd tangent
vectors as well as even ones and the Virasoro action on $\pgs gs$ is
extended to an action of the Neveu-Schwarz algebra on $\ph gs$.
Following the conventions of \EFEFS, we take the generators of this
algebra to be
\eqn\nsa{\eqalign{
   l_n& \leftrightarrow -z^{n+1}\dbd{z}-\half(n+1)z^n\theta\dbd{\theta}\cr
     g_k& \leftrightarrow \half z^{k+\ha}
          \bigl(\dbd{\theta} -\theta\dbd{z}\bigr).}}
The factor of 1/2 in the definition of the $g_k$ is conventional and
it will show up later. Again, we associate states in a Hilbert space to
a triple of a surface with a puncture and a superconformal coordinate at
the puncture. The action of the generators in \nsa corresponds to an action
on the Hilbert space of states by
\eqn\NSA{\bra{\Sigma, z-\epsilon z^{n+1}+\half\hat{\epsilon}\theta
z^{k+\ha}}=\bra{\Sigma, z}\bigl(1+\epsilon L_n +\hat{\epsilon}
G_k\bigr)+\cdots .}

The measure, on $\ph gs$, is given by (again, in abbreviated form)
\eqn\hmeas{\eqalign{\Oht(\vt1,\vt2,\ldots,\vt{3g-3+s},
     \Ut1,\Ut2,\ldots,\Ut{2g-2+s})&=\cr
 \bra{\Sigma,(z_1,\th_1),\ldots}\Bv1\ldots\Bv{3g-3+s}
           \dBnu1\ldots&\dBnu{2g-2+s}
         \ket{\psi_1}_{P_1}\otimes \cdots\otimes\ket{\psi_s}_{P_s}.}}
Here $B(z,\theta)=b(z)+\theta \beta(z)$ and $v(z,\theta)=v_1(z)+\theta
v_2(z)$
($\nu(z,\theta)=\nu_1(z)+\theta\nu_2(z)$) is an even (odd) vector field.
The notation $B[v]$ corresponds to
$$ \oint \dd z \dd\theta B(z,\theta)v(z,\theta). $$ A similar expression
holds for $B[\nu]$. See \EFEFS and \AGNSV for more details.
The pulled back measure on $\mh gs$ is exactly analogous to the bosonic
one given in \measm.  It is shown in \EFEFS that this measure changes
only by a total derivative for different choices of
$\s$ if the states that are inserted obey
the WPSC, \wps. Thus, integrated answers are independent of the choices
made. In addition, it is shown that $Q$ again acts as the exterior
derivative on $\ph gs$.

As in the bosonic case, this general formalism is easily adapted to the
computation of heterotic dilaton contact terms. Once again contact terms
are best dealt with by making a superconformal transformation from the
picture of the two operators colliding to one where they are pinched off
from the rest of the surface. There are a few differences here though. We
have to use a plumbing fixture suitable for the heterotic string and,
more importantly, there is an odd modulus associated to the
three-punctured super-sphere that has to be considered.

To begin with, we look at the three-punctured super-sphere with
coordinates $w$ and $\xi$. The standard (bosonic) three-punctured sphere
is rigid, i.e., it has no moduli associated with it. This is because
we can use $SL(2,{\bf C})$ invariance to carry our three marked points into
three standard points, say 0, 1, and $\infty$. On the super-sphere we
have $Osp(2,1)$ invariance and this can be used to fix the bosonic
coordinates of our three marked points to be 0, 1, and $\infty$.
However, we can only fix two of the three fermionic coordinates.
This means that the third unfixed fermionic coordinate is a modulus. The
dimension of the moduli space, $\mh 03$,  is thus $(0|1)$.
Using $Osp(2,1)$ freedom, we can locate our three points at (0,0),
(1,$\tau$) and ($\infty$,0). $\tau$ is the leftover odd modulus of the
three-punctured super-sphere. Our standard plumbing fixture will be to
locate the point $P$ at 1 and $Q$ at 0 and sew the point at $\infty$ onto
the rest of the surface. Good coordinates at $\infty$ are simply $-1/w$
 and $\xi/w$. This is will make the sewing simple.
Since the states will be inserted at $P$ and $Q$, we should use the
most general superconformal holomorphic coordinates the vanish at 1 and 0.
At 0, we use the coordinates
\eqn\zercor{w+\aa_1 \tau w \xi + \aa_2 w^2 +\aa_3 \tau w^2 \xi +\cdots}
(we give just the even coordinate) and at
(1,$\tau$) we use
\eqn\onecor{(w-1+ \tau \xi)+  a_1  \tau (w-1+ \tau \xi) (\xi - \tau)
+a_2 (w-1+ \tau \xi)^2 +  a_3 \tau (w-1+ \tau \xi)^2 (\xi -
\tau)+\cdots \; .}

The coefficients $\aa_i$ and $a_i$ for $i\ge 3$ turn out not to
affect the dilaton calculations.

This fixture is attached to the rest of the surface by using
the standard sewing prescription. If $(x,\xi')$ and $(y,\psi)$ are the
coordinates of the two regions of ${\bf C}^{1|1}$ that are being sewn
together, then they are joined by
\eqn\sewing{x y=-t^2\,, \quad x\psi=-t\xi'\,,\quad y\xi'=t \psi\,,}
where $t$ is the sewing modulus. Notice it is $t^2$ that plays the role
that $q$ played in the bosonic case. Also, the counting of the moduli
works out: $P$ and $Q$ each have an even and an odd modulus and they are
replaced by the moduli of the attachment point (which we again take to
be $Q$) and the one even modulus of the sewing and the odd modulus of
the three-punctured super-sphere,$\,\tau$.

Ideally, one would like to do the dilaton two-point function on the
sphere in a way analogous to the bosonic case. However, it turns out
to be too tedious and complicated. Thus, we will proceed by carrying out
a local, one patch calculation, and assume that a `good' coordinate
system similar to the one constructed in Section 4 could also be
constructed for the heterotic case.
One shortcoming of this approach is that we will miss total derivatives
with respect to the moduli of $Q$, preventing us from easily calculating
everything that we would like. However, all is not lost and we will
still be able to
show that the tachyon divergence is absent. Moreover, by using
arguments involving the decoupling of genuine BRST-exact states (which we
know to be true on general grounds), we will be able to demonstrate that
the dilaton equation works in the heterotic string, even when more than
one dilaton is in the correlation function.

In the bosonic case, the dilaton was given in \bosdil.
Similarly, for the heterotic string the dilaton is given by \EFEFS
\eqn\hetdil{\ket{D}=2 (Q+\bar{Q})(c_0-\bar{c}_0) \delta(\g_{1/2})\ket{0}}
It is convenient to work with $\ket{D_1}$ and $\ket{D_2}$, defined in
\EFEFS, instead, where
\eqn\dildef{\eqalign{
  \ket{D_1}&=Qc_0 \delta(\g_{1/2})\ket{0}=\dilone\cr
  \ket{D_2}&=\bar{Q}\bar{c}_0 \delta(\g_{1/2})\ket{0}=\diltwo.}}
Then $\ket{D}$  is simply
\foot{The factor of 2 is conventional and is
chosen so that the one-point function of  $\ket{D}$ on the sphere is
normalized to $4\pi i Z=-2\pi i \chi Z$ as in \TCCT. This differs from the
conventions in \EFEFS.}
$$ \ket{D}=2(\Done+\Dtwo).$$

It was shown in \EFEFS that the orthogonal combination $\Done-\Dtwo$ is a
global total derivative and decouples from all correlation functions.
This is exactly analogous to situation described earlier in the
discussion of the bosonic string dilaton.
Once again this state is $Q$ of something that obeys the WPSC and
hence gives rise to a globally defined total derivative. Thus,
one can work equally well
with $\ket{D}$, $4\Done$, or $4\Dtwo$ since they all differ by adding in
multiples of $\Done-\Dtwo$ which contributes a genuine global
total derivative. Notice that $\ket{D_1}$ and $\ket{D_2}$ are distinct
operators whereas in the bosonic string the corresponding states were
merely the barred and un-barred versions of the same state, $2c_1
c_{-1}\ket{0}$ and $2\bar{c}_1 \bar{c}_{-1}\ket{0}$. In the
bosonic case, it was the contact interaction between the holomorphic and
antiholomorphic pieces of the dilaton that produced the tachyonic
divergence \TCCT. In \DESRST it was seen that this can be avoided by
working with the purely holomorphic dilaton only. From this one might
expect that the tachyon divergence would appear in the $\Done\Dtwo$ contact
term. This is
precisely the term that will be easiest to calculate (as well as
$\Dtwo\Done$), and, indeed divergent terms appear. However,
they will be killed by the integration over the odd moduli and there is
no tachyon divergence.

It is
somewhat unfortunate that the local representation of the curvature used
below turns
out to obscure the total derivatives in the moduli of the
point $Q$ appearing in the $\Done\Done$ or $\Dtwo\Dtwo$ contact terms
which are known to be present on general grounds.
In particular, it is found that the $\Done\Done$ and $\Dtwo\Dtwo$ terms
depend on the higher curvature coefficients in the local expansion of the
coordinates introduced below, while the $\Done\Dtwo$ and $\Dtwo\Done$
terms do not. But this is not a serious hinderance since we can use the
decoupling of $\Done -\Dtwo$ to our advantage, as will be seen below.

\subsec{Coordinates}
With these preliminaries out of the way, we can now adapt the formalism
to our specific calculation.
We work in one patch where the coordinates interpolate from those
appropriate to the two states far apart to those appropriate for them
approaching each other. Things will be simplest if we put the curvature
around only one of the points ($Q$ here). Then integrating over the
position of $P$ will merely give us only the contact term since the
dilaton doesn't couple to flat backgrounds. So, using the coordinates
$(z,\th)$ for the worldsheet and letting $(\rr,\rrho)$ and $(r,\rho)$
be the positions of $P$ and $Q$, respectively, we choose
coordinates~\eqn\Pcor{\zeta_P(\cdot)=z(\cdot)-\rr+\rrho \th(\cdot)}~at
$P$ and
\eqn\Qcor{\eqalign{\zeta_Q(\cdot)=&\bigl(z(\cdot)-r+\rho \th(\cdot)\bigr) +
          \rho\rb R_1 \bigl(z(\cdot)-r+\rho\th(\cdot)\bigr)
               \bigl(\th(\cdot)-\rho\bigr)+\cr
            &\rb R_2 \bigl(z(\cdot)-r+\rho \th(\cdot)\bigr)^2 +
          \rho\rb R_3 \bigl(z(\cdot)-r+\rho\th(\cdot)\bigr)^2
               \bigl(\th(\cdot)-\rho\bigr)+\cr
            &\rb R_4 \bigl(z(\cdot)-r+\rho \th(\cdot)\bigr)^3 +
\cdots}} at $Q$,
where the
coefficients $R_i$  make our coordinate slice non-holomorphic.
What we have given is a small $r$ expansion of the coordinates in a
curved background in which the
mixing of holomorphic and antiholomorphic coordinates is akin to the
effect of curvature in Polchinski's scheme \JPvert.
This is our local curvature picture, and the fact that we have done a
small $r$ expansion will obscure some total derivatives that we know are
there on general grounds.
In the next section, we
will find that $R_1$ and $R_2$ are related to the scalar curvature
by demanding that the one-point function $\vev{(D_1-D_2)}$
vanish on the sphere and that the one-point function of
$\vev{(D_1+D_2)}$ be properly normalized. The higher curvature
coefficients cannot be determined in this way. And, as stated above, it
turns out that only the $\Done\Dtwo$ and $\Dtwo\Done$ terms are
independent of the $R_i, i\ge3$.

When the points are close together, coordinates should go over to those
given by the plumbing fixture. The sewing in \sewing gives
(see \tfig\hetplumb)
\eqn\sew{w={\zeta_Q\over t^2}\,,\quad \xi=-{\check{\zeta_Q} \over t}.}
Substituting this into the coordinates given in \zercor and \onecor
results in
\eqn\sewncor{\eqalign{\phi_Q(\cdot)=&{\zeta_Q(\cdot)\over t^2}+
                   \aa_1\tau {\zeta_Q(\cdot) \check{\zeta_Q}(\cdot) \over t^3}+
                         \aa_2 {\zeta_Q(\cdot)^2 \over t^4} +\cdots\cr
\phi_P(\cdot)=&\Sigma(\cdot) +a_1 \tau\Sigma(\cdot) \check\Sigma(\cdot)
                  +a_2 \Sigma(\cdot)^2 +\cdots\,,}}
with
\eqn\Sigdef{\eqalign{
    \Sigma(\cdot)=&\bigl({\zeta_Q(\cdot) \over t^2} -1 -
                {\tau \check{\zeta_Q}(\cdot) \over t}\bigr)\cr
\check\Sigma(\cdot)=&\bigl({\check{\zeta_Q}(\cdot)\over t}+\tau\bigr).}}
Also, there is a sign that has been absorbed by redefining the
arbitrary odd-indexed $a_i$'s and $\aa_i$'s.

\insfig\hetplumb{Heterotic plumbing fixture with coordinates. The
construction is similar to that in \bplumb, but now there is an odd
modulus, $\tau$, associated with the three-punctured super-sphere.}{2.5}
{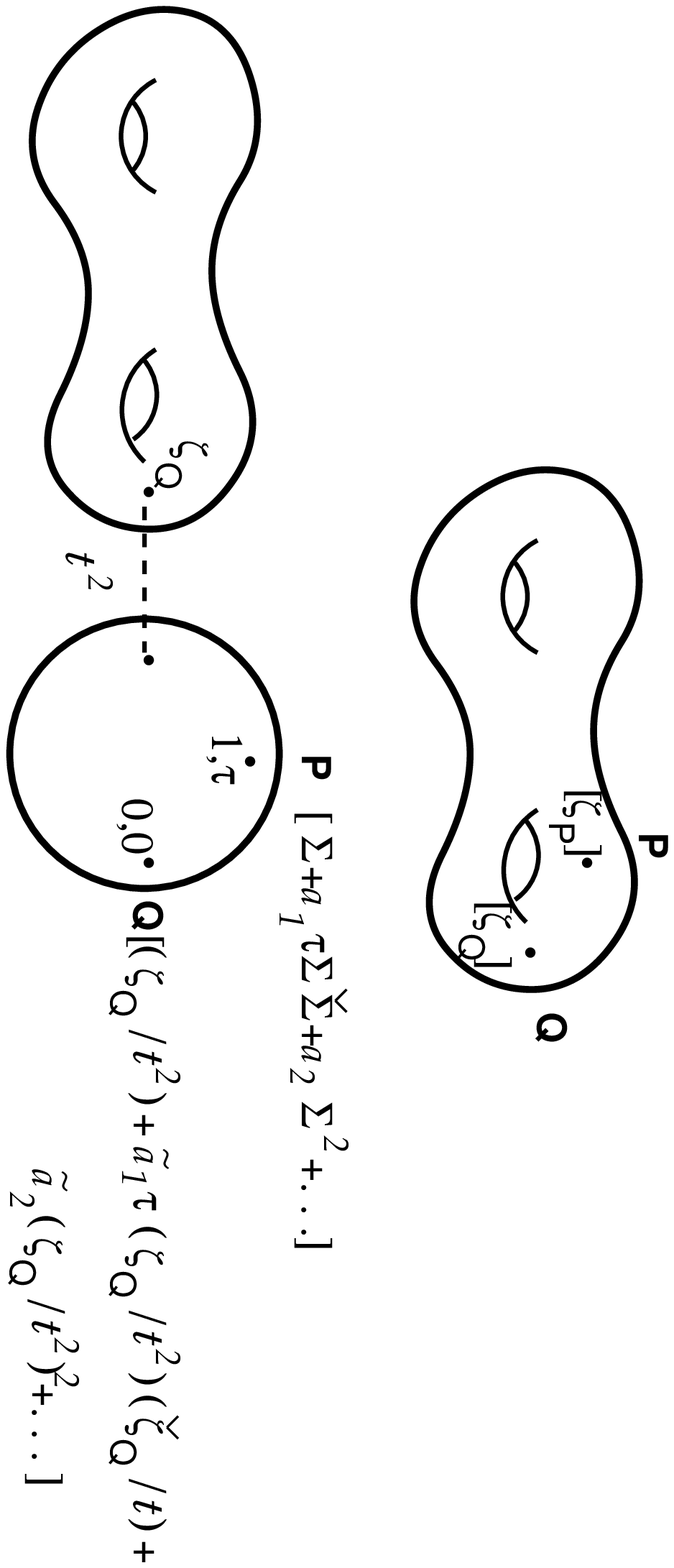}{90}{55}{55}{455}{-90}

Since we are just trying to pick out the contact term in this
calculation, we will take $r$,$\rho$,$t$,and $\tau$ to be the moduli.
Thus we have to eliminate $\rr$ and $\rrho$ in favor of $t$ and $\tau$
in $\zeta_P$. This is done by again demanding that $\phi_P(P)=0$ or that
$\zeta_Q(P)=t^2$ and $\check{\zeta_Q}(P)=-\tau t$. In addition, the contact
terms will only be proportional to the curvature in the lowest order so
we will only keep terms that are linear in $R_1$ and $R_2$. We find that
\eqn\rrexp{\eqalign{\rr=&r+t^2 +t \rho \tau+\rb R_1 t^3 \rho\tau-
                    \rb R_2 t^3\rho\tau -\rb R_2 t^4+\cdots \cr
        \rrho=&\rho-\tau t+\rb R_1 \rho t^2+\rb R_2 \tau t^3+\cdots\,.}}
With the above definitions and relations in mind, the coordinate slice
for our calculation will just be the linear interpolation between the
coordinate regions in \Pcor,\Qcor, and \sewncor. The interpolation
function is $f(|t|)$ and goes smoothly from 0 to 1 as $|t|$ goes
from 0 to $\infty$. Thus,

\eqn\slice{\eqalign{\sigma_P(\cdot)=&{f(|t|) \over t^2}
                 \zeta_P(\cdot)+(1-f(|t|)) \phi_P(\cdot)\cr
                 \sigma_Q(\cdot)=&{f(|t|) \over t^2}\zeta_Q(\cdot) +
                    (1-f(|t|)) \phi_Q(\cdot)\,.}}
The reader is reminded that there are odd coordinate functions that go
along with these expressions that can be found using \sci-\sciii.
One final note: We have been displaying the coordinates for the
holomorphic sector. The appropriate expressions for the antiholomorphic
sector are obtained by simply setting all odd parameters to zero in the
holomorphic expressions and complex-conjugating the resulting expressions.

\subsec{The Measure}
With coordinates in hand, we can now calculate the $b$-insertions that are
required to produce the measure. The method is identical to that in the
bosonic case, with appropriate generalizations. The pushforwards are
given by
\eqn\hetpush{
  \s_{*}({\dbd m})={\der {\s}{m}} {\dbd {\s}} +
                  {\der {\bar \s}{m}} {\dbd{\bar \s}}+
                  {\der {\sth}{m}} {\dbd{\sth}}
}
where $m$ is one of $r\,$,$\,\rb\,$,$\,\rho\,$,$\,t\,$,$\,\tb\,$,
or $\,\tau\,$. There
is no $\bar {\sth}$ term since we are doing the heterotic string.
The corresponding operator insertions that form the measure are found
by folding pushforwards with the $B$'s. Since the resulting insertions are
quite long, we have displayed them in Appendix B. In addition, the
calculation is done at $r=\rb=0$ for convenience. Thus the pushforwards
are calculated by first differentiating and then setting $r$ and $\rb$
to 0. Finally, the insertions
appropriate for each case of interest can be identified and then picked
out of the measure\foot{There are many terms that contribute and
$Mathematica$ was quite useful here as well.}. The insertions for the
various states are
\eqn\hetins{\eqalign{
b_{-1}\bar{b}_{-1}\db{-{1\over2}}&\quad{\rm for}\quad\ket{\psi}\cr
b_{-1}\be{1\over2}\db{-{1\over2}}&\quad{\rm for}\quad\Done\cr
b_{-1}\bar{b}_{1}\bar{b}_{-1}\dbp{-{1\over2}}&\quad{\rm for}\quad\Dtwo}.}
These are the only combinations of ghost insertions that contribute.
(Although later we will see that there are terms with additional
derivatives of delta-functions that also contribute. These additional
contributions are formally equivalent to the above insertions.)
It should be noted that
for $\ket{\psi}\,$ (a SPS), the integration over the odd moduli would
give zero if it were not for the fact that the state $\bra{\Sigma}$
itself depends on the odd moduli. In fact, using \NSA,
(a similar equation applies to the coordinates at $Q$ without the term
 proportional to $\tau$),
\eqn\state{\bra{\Sigma,\sigma_P}=
              \bra{\Sigma,\tilde{\sigma}_{P_0}}\left(1+2 \rho
G^{P}_{-{1\over2}} - 2 t \tau  G^{P}_{-{1\over2}} + \cdots\right)
  t^{2 L_0} \bar{t}^{2 \bar{L}_0},}
where $\tilde{\sigma}_{P_0}$ is $\sigma_P$ with $\rho$ and $\tau$ set to
zero and with an overall factor of $t^2$ removed. This is because the
coordinates in \slice and \Pcor and \Qcor differ by an overall factor of
$t^2$. These factors of $t^{2 L_0}$ matter for a SPS because the state
$$ \b{-1}{Q}\bb{-1}{Q}\dbt{\mha}{Q}\ket{\psi} $$ is not
annihilated by $L_0$ and in fact has $L_0=1/2$ and $\bar{L}_0=1$.
The dilaton, on the other hand, is annihilated by the $L_0$ and $\bar{L}_0$
and no such scale factors are necessary. Furthermore,
the $G_{-{1\over2}}$ combines with the $\db{\mha}$ to give a picture
changing operator when inserting a SPS. However, there are no
contributions involving the picture-changing operator for the dilatons
\EFEFS. Finally, the expansion of the state was done at $r=\rb=0$.

As an example, let's calculate the insertions for $\ket{\psi}$,$\ket{D_1}$,
 and
$\ket{D_2}$ in the background given in \Qcor. We find (renaming $\zeta_Q$,
$\sigma$)
$$\eqalign{\sigma_*[\dbd{r}]&=-1 + R_2 {\bar \s}^2 +\cdots\cr
       \sigma_*[\dbd{\rb}]&=R_2 \s^2 +R_1 \rho \s \sth -1 +\cdots\cr
       \sigma_*[\dbd{\rho}]&=\rho + \sth +\cdots\; ,}$$
where the -1 in the $\rb$ pushforward is from the barred part and the
terms from derivatives of $\sth$ have been omitted (they always just
give the second half of the superconformal vector fields and we can read
off the insertions without considering them). These give rise to
\eqn\qcorp{\eqalign{B[\s_*[\dbd{r}]]&=-b_{-1}+R_2 \bar{b}_{1}+\cdots \cr
 B[\s_*[\dbd{\rb}]]&=R_2 b_{1}-2 R_1 \rho \be{\ha}-\bar{b}_{-1}+\cdots\cr
    B[\s_*[\dbd{\rho}]]&=-\rho b_{-1}+2 \be{\mha}+\cdots\;, }}
where we have kept only the contributing terms. Picking off the measure
contribution for $\Done$ and $\Dtwo$ (see \hetins), we have
\eqn\doneq{\eqalign{(-b_{-1})& (-2 \rho R_1 \be{\ha})\dg{2 \be{\mha}}\Done\cr
&=-\rho R_1 b_{-1} \be{\ha}\dg{\be{\mha}}\Done\cr
&={1\over 2} \rho R_1 b_{-1} \be{\ha}\db{\mha} c_1 \g_{\mha}
                          \dg{\g_{\ha}}\ket{0}\cr
&= {1\over2} \rho R_1 \ket{0},\cr}}
and
\eqn\dtwoq{\eqalign{
(R_2 \bar{b}_1)&(-\bar{b}_{-1})\dg{-\rho b_{-1}+2 \be{\mha}}\Dtwo\cr
&={1\over4}R_2 \bbc{1}\bbc{-1}\rho b_{-1} \dbp{\mha}\Dtwo\cr
&=-{1\over2}\rho R_2 \bbc{1}\bbc{-1}b_{-1}\dbp{\mha}c_{1}\cb{1}
\cb{-1} \be{\mha}\dg{\g_{\ha}}\ket{0}\cr
&=-{1\over2} \rho R_2 \ket{0}.}}
Thus, $R_1=-R_2$ will enforce the decoupling of $(\Done-\Dtwo)$ which
we know to be true on general grounds \EFEFS. Furthermore, $R_1$ and
$R_2$ are proportional to the scalar curvature since $2(\Done+\Dtwo)$
is the dilaton.

In addition, we will need the insertions for a
SPS in this background. We have, from \hetins and \qcorp,
\eqn\spsins{\eqalign{(-b_{-1})&(-\bbc{-1})\dg{2 \be{\mha}}\ket{\psi}\cr
 &={1 \over 2}b_{-1}\bbc{-1}\db{\mha}\ket{\psi}.}}
The conventional factor of 2 may seem out of place,
but it will be cancelled by a similar factor of 2 in the expansion of
the state $\bra{\Sigma,\zeta_Q}$ when the $\rho$ integral is done.
These factors of 2 appear because of our convention in defining the
$g_k$ in \nsa.

\newsec{The Dilaton Equation}
All the ingredients have now been assembled to demonstrate the dilaton
equation in the heterotic string. We are to compare
\eqn\dilexp{\eqalign{\int \tmeas&
 \rmeas\bra{\Sigma,\sigma_P,\sigma_Q}\cdots
   \B{\dbd{t}}\B{\dbd{\tb}}\delta\big[\B{\dbd{\tau}}\big]\cr
     &\times\B{\dbd{r}}\B{\dbd{\rb}}\delta\big[\B{\dbd{\rho}}\big]
\kpkq{\Phi_1}{\Phi_2}}}
with the state gotten by integrating over $t$,$\tb$, and $\tau$,
\eqn\dilexpi{\int \rmeas \bra{\Sigma,\zeta_Q}\cdots
     B[\zeta_*({\dbd{r}})]B[\zeta_*({\dbd{\rb}})]
\delta[B[\zeta_*({\dbd{\rho}})]] \ket{\Phi_2}^Q.}
Physically this corresponds to integrating over the position of the
operator at $P$.
The dots represent the insertions for the moduli not associated to
the locations of the points $P$ and $Q$ and the $\Phi_i$ will be one of
$\ket{\psi}$(a SPS), $\Done$, or $\Dtwo$. Actually, of the eight
possibilities involving at least one dilaton, only four can be easily
obtained using the way that we have represented the curvature on the surface.
These are $\kpkq{D_1}{\psi}$,
$\kpkq{D_2}{\psi}$, $\kpkq{D_1}{D_2}$, and $\kpkq{D_2}{D_1}$. They only
depend on the coefficients $R_1$ and $R_2$ in \Qcor. The other contact
terms all depend on the higher $R_i$, a signal that in expanding around
$r=0$, we have made certain total derivatives in $r$ and $\rb$ difficult
to see. It should be noted that the same problem occurs in the bosonic
string if one tries to do that calculation in the same way.
We consider each of the four terms that we can easily calculate
in turn.

\subsec{Dilaton-Strong Physical State}
As in the bosonic case, the $\kpkq{D_1}{\psi}$ and
$\kpkq{D_2}{\psi}$  calculations determine the values of $a_1$ and
$a_2$ that are needed to make the dilaton equation work for correlation
functions with one dilaton and the rest SPS's. This is exactly analogous
to the calculation done in the background in \Qcor, which determined the
relationship between $R_1$ and $R_2$, but now the insertions appropriate
for the coordinates in \slice are used. The corresponding ghost
insertions are given in Appendix B.

The insertions that reduce $\kpkq{D_1}{\psi}$ to $\kpkq{0}{\psi}$ are
simply those given in \hetins,
$$ \b{-1}{P} \bet{\ha}{P} \dbt{\mha}{P}
       \b{-1}{Q} \bb{-1}{Q}\dbt{\mha}{Q}.$$
Thus, \dilexp becomes
\eqn\dones{\eqalign{\int &\tmeas \rmeas\cr
&\times \bra{\Sigma,\s_P,\s_Q}\left(-{a_1 \tau t f'(|t|)\over
     2 |t|^5}\right) \b{-1}{P} \bet{\ha}{P} \dbt{\mha}{P}
       \b{-1}{Q} \bb{-1}{Q}\dbt{\mha}{Q} \kpkq{D_1}{\psi}.
}}
This expression is to be integrated over $t$, $\tb$, and $\tau$,
leaving the SPS inserted at $Q$ with its ghost insertions. To show the
dilaton equation, this state is to be compared with the one
corresponding to the SPS inserted with the coordinate $\zeta_Q$ in
\Qcor. Recalling \state and the discussion following it,
and using \dildef, \dones becomes
\eqn\donesa{\eqalign{\int & \tmeas \rmeas
\bra{\Sigma,\tilde{\sigma}_{Q_0}}\bigl(1+2\rho G^{(Q)}_{\mha}\bigr)\cdots\cr
&\times\left({a_1 \tau  f'(|t|)\over
     2 |t|}\right) \ha \b{-1}{P} \bet{\ha}{P} \dbt{\mha}{P}
 \b{-1}{Q} \bb{-1}{Q}\dbt{\mha}{Q}\dilonep{P}\otimes\ket{\psi}^Q\cr
}}
where the factor of $1/2$ is picked out so that the measure for the SPS
insertion agrees with that given in \spsins, and the dots represent
suppressed ghost insertions. The $b$-insertions at $P$ reduce the
dilaton to the vacuum, which can be erased.  An additional subtlety is
that we have to
move $\rho\tau$ to the front of the expression so that we can integrate over
$\tau$. Thus we move $\rho G^Q_{\mha}$ through the ghost insertions,
move the $\tau$ through the $G^Q_{\mha}$ picking up a minus sign and
then move the $\rho \tau$ back through the ghost insertions. This
results in
\eqn\donesb{
-\int  \tmeas \rmeas \rho\tau\bra{\Sigma,\tilde{\sigma}_{Q_0}}
\left({a_1 f'(|t|)\over 2 |t|}\right)
 \ha (2 G^{(Q)}_{\mha})\b{-1}{Q} \bb{-1}{Q}\dbt{\mha}{Q}\ket{\psi}^Q.
}
The insertions at $Q$ are the same as the ones that would result from
using the coordinate $\zeta_Q$. Furthermore, we can integrate over
$\tau$ and do the $t$ integral, recalling that
 $\dd t\we \dd \tb=-2 i|t|\dd |t|\we\dd \theta$. The result is

\eqn\donesi{2 \pi i a_1 \int\rmeas \bra{\Sigma,\zeta_Q}
\ha\b{-1}{Q}\bb{-1}{Q}\dbt{\mha}{Q}\ket{\psi}^Q.}

Similarly, $\kpkq{D_2}{\psi}$ gives
\eqn\dtwos{\eqalign{
\int &\tmeas \rmeas\cr
&\times \bra{\Sigma,\s_P,\s_Q}\left(-{a_2 \tau t f'(|t|)\over
     8 |t|^5}\right) \b{-1}{P} \bb{1}{P}\bb{-1}{P} \dbpt{\mha}{P}
       \b{-1}{Q} \bb{-1}{Q}\dbt{\mha}{Q} \kpkq{D_2}{\psi}\cr
=&\int \tmeas \rmeas\bra{\Sigma,\tilde{\s}_Q}
 \left({a_2 \tau f'(|t|)\over 2 |t|}\right) \cr
     &\times\ha\b{-1}{P} \bb{1}{P}\bb{-1}{P} \dbpt{\mha}{P}
       \b{-1}{Q} \bb{-1}{Q}\dbt{\mha}{Q} \diltwop{P}\otimes\ket{\psi}^Q\cr
=&-\int \tmeas \rmeas\rho\tau\bra{\Sigma,\tilde{\s}_{Q_0}}
 \left({a_2 \tau f'(|t|)\over 2 |t|}\right)
  \ha(2 G^{(Q)}_{\mha})\b{-1}{Q} \bb{-1}{Q}\dbt{\mha}{Q}\ket{\psi}^Q\cr
=& 2 \pi i a_2 \int\rmeas \bra{\Sigma,\zeta_Q}\ha\b{-1}{Q}\bb{-1}{Q}
               \dbt{\mha}{Q}\ket{\psi}^Q.
}}
Recalling that the dilaton $\ket{D}=2(\Done +\Dtwo)$, the state
$4 \Done$, and the state $4 \Dtwo$ are all equivalent in correlation
functions,
we see that the choice $a_1=a_2=-1/4$ will give the correct
normalization for the dilaton contact term in \dil, at least when all of
the other operators are SPS's. The final piece is
to check that the dilaton-dilaton contact terms behave appropriately.
This is done in the following section.

\subsec{Dilaton-Dilaton}
We now turn to $\kpkq{D_1}{D_2}$ and
$\kpkq{D_2}{D_1}$. Both calculations are presented so that we can check
that our answer is consistent, i.e., that it does not matter which
dilaton we put at $P$ and which we put at $Q$. Also, it allows us to
explicitly check the decoupling of $\ket{D_1}-\ket{D_2}$. The calculations are
almost identical to those of the last section, except that there are now
many more ghost insertions that contribute and each can be gotten by
selecting the insertions from the pushforwards in several different ways.

There are five different combinations of ghost insertions that appear in
the measure that reduce $\kpkq{D_1}{D_2}$ to $\kpkq{0}{0}$:
$$ \eqalign{&\b{-1}{P} \bet{\ha}{P} \dbt{\mha}{P}
     \b{-1}{Q}\bb{1}{Q}\bb{-1}{Q}\dbpt{\mha}{Q}\cr
&\b{-1}{P} \bet{\ha}{P}\dbt{\mha}{P}\b{-1}{Q}
    \bb{1}{Q}\bb{-1}{Q}\bet{\mha}{Q}\dbppt{\mha}{Q}\cr
&\b{-1}{P} \bet{\ha}{P}\bet{\mha}{P} \dbpt{\mha}{P}\b{-1}{Q}
    \bb{1}{Q}\bb{-1}{Q}\dbpt{\mha}{Q}\cr
&\b{-1}{P} \bet{\ha}{P}\bet{\mha}{P}\bet{\mha}{P} \dbppt{\mha}{P}\b{-1}{Q}
    \bb{1}{Q}\bb{-1}{Q}\dbpt{\mha}{Q}\cr
&\b{-1}{P} \bet{\ha}{P}\bet{\mha}{P} \dbpt{\mha}{P}\b{-1}{Q}
    \bb{1}{Q}\bb{-1}{Q}\bet{\mha}{Q}\dbppt{\mha}{Q}.\cr
}$$
All other possible ghost insertions do not appear for the coordinate
system that we are using. Each of the above combinations also occurs
several times in the expansion of the measure. After carefully
accounting for all relative minus signs for ordering and moving
$\rho \tau$ to the front, and using the formal rules
$x \delta'(x)=-\delta(x)$ and $ x^2 \delta''(x)=2 \delta(x)$
\foot{These formal rules can be derived from a more rigorous point of
view, starting from an axiomatic set of definitions for $\delta(\beta)$ and
$\delta(\gamma)$ and how they act\DUNP.}, we find
that \dilexp becomes
$$\eqalign{\int& \tmeas \rmeas \bra{\Sigma,\s_Q}\cr
&\times \left({\rho \tau  a_1 R_2 f'(|t|)\over 4 |t|}+\cdots\right)
\b{-1}{P} \bet{\ha}{P} \dbt{\mha}{P}
 \b{-1}{Q}\bb{1}{Q}\bb{-1}{Q}\dbpt{\mha}{Q} \kpkq{D_1}{D_2},}$$
where the dots represent divergent terms that do not contain $\rho$ and
$\tau$ and so are killed by the integration over the odd moduli. (There
are terms in the expansion of the state $\bra{\Sigma}$ that are proportional
to $\rho\tau L_n,\quad n\ge -1$, but these annihilate the vacuum and so
they do not contribute.) Putting in the state $\Done$ and simplifying
gives
$$\eqalign{&\int \tmeas \rmeas \bra{\Sigma,\s_Q}\cr
&\left(-{\rho \tau  a_1 R_2 f'(|t|)\over 8 |t|}\right)
\b{-1}{P} \bet{\ha}{P} \dbt{\mha}{P}\b{-1}{Q}\bb{1}{Q}\bb{-1}{Q}
\dbpt{\mha}{Q}\dilonep{P}\otimes\ket{D_2}^{Q}\cr
=&\int \tmeas \rmeas \bra{\Sigma,\s_Q}
\left(-{\rho \tau  a_1 f'(|t|)\over 2 |t|}\right)
{R_2 \over 4}\b{-1}{Q}\bb{1}{Q}\bb{-1}{Q}
\dbpt{\mha}{Q}\ket{0}^P\otimes\ket{D_2}^{Q},}$$
where, again, the factor of $R_2/ 4$ is chosen to agree
with the insertions for $\Dtwo$ in \dtwoq.
Performing the intergrations over $t$ and $\tau$ is now straightforward,
with the result
\eqn\doteq{2 \pi i a_1 \int \rmeas \bra{\Sigma}{\rho R_2\over4}
 \b{-1}{Q}\bb{1}{Q}\bb{-1}{Q}\dbpt{\mha}{Q}\ket{D_2}^Q.}
This is exactly the answer that we wanted for the dilaton equation.
Comparing this result to \donesi, we see that the contact term between
two dilatons is indeed normalized properly.

Similarly, there are three combinations for $\kpkq{D_2}{D_1}$:
$$\eqalign{&\b{-1}{P}\bb{1}{P}\bb{-1}{P}\dbpt{\mha}{P}
     \b{-1}{Q} \bet{\ha}{Q}\dbt{\mha}{Q}\cr
&\b{-1}{P}\bb{1}{P}\bb{-1}{P}\bet{\mha}{P}\dbppt{\mha}{P}
     \b{-1}{Q} \bet{\ha}{Q}\dbt{\mha}{Q}\cr
&\b{-1}{P}\bb{1}{P}\bb{-1}{P}\dbpt{\mha}{P}
     \b{-1}{Q} \bet{\ha}{Q}\bet{\mha}{Q}\dbpt{\mha}{Q}.}$$
The result here is
$$\eqalign{
\int &\tmeas\rmeas\bra{\Sigma,\s_Q}
\left(-{\rho \tau a_2 R_1 f'(|t|)\over 4 |t|}+\cdots\right)\cr
&\quad\times\b{-1}{P}\bb{1}{P}\bb{-1}{P}\dbpt{\mha}{P}
\b{-1}{Q} \bet{\ha}{Q}\dbt{\mha}{Q}
  \kpkq{D_2}{D_1}\cr
=&\int \tmeas\rmeas\bra{\Sigma,\s_Q}
\left({\rho \tau a_2 R_1 f'(|t|)\over 2 |t|}\right)\cr
&\quad\times\b{-1}{P}\bb{1}{P}\bb{-1}{P}\dbpt{\mha}{P}
\b{-1}{Q} \bet{\ha}{Q}\dbt{\mha}{Q}
 \diltwop{P}\otimes \ket{D_1}^{Q}\cr
=&\int \tmeas\rmeas\bra{\Sigma,\s_Q}
\left(-{\rho \tau a_2 f'(|t|)\over 2 |t|}\right)
\left(-R_1\right)\b{-1}{Q} \bet{\ha}{Q}\dbt{\mha}{Q} \ket{D_1}^{Q}
,}$$
and, after integrating,
\eqn\dtoeq{2 \pi i a_2 \int\rmeas \bra{\Sigma}(-R_1)
      \b{-1}{Q} \bet{\ha}{Q}\dbt{\mha}{Q}\ket{D_1}^Q.}
Comparing this to \dtwos, we again see that the dilaton-dilaton contact
terms are properly normalized. Thus, the dilaton equation holds in the
heterotic string, even when other dilatons are in the correlation
functions.

\newsec{Conclusions}

Dilatons and contact terms have had a long and sometimes confusing history,
and while this work has shed some light on old issues, it also raises some
interesting new points worthy of further investigation.
In this paper we have further examined the properties of dilaton
contact terms in string theories. In the case of the bosonic strings we
sketched how to construct `good' coordinate families for doing the kinds
of calculations contained in this paper on higher genus surfaces. This
addresses some of the tacit assumptions in \TCCT.
In addition, we are left with the fact that the dilaton equation in the
bosonic string fails when there is more than one dilaton in a
correlation function. It is not clear what the origin of this failure is,
but one could speculate the it is due to the tachyon. This is also
somewhat dismaying since it implies that while the insertion of one
dilaton in a correlation function corresponds to the first variation with
respect to the string coupling constant, the insertion of two dilatons
is not just the second variation of the string coupling constant.
Although we do not take this point up here, we feel that this merits
further examination.

In the heterotic case we have seen the absence of the tachyonic
divergence and that the dilaton equation is valid. This would seem
to strengthen the idea that the failure of the dilaton equation in the
bosonic string is in some way due to the tachyon. Thus it would seem
that the dilaton can really be viewed as the operator which corresponds
to varying the string coupling constant in the heterotic string. In addition,
although we have not done so here, it would be straightforward to extend
the method for constructing a `good' coordinate family in the bosonic
string to the heterotic case.

\bigskip
\centerline{Acknowledgements}
It is a pleasure to thank Jacques Distler for his patient explanations
and guidance, and to acknowledge helpful conversations with Phil Nelson,
Eugene Wong, and Suresh Govindarajan.
And many thanks to the Outlaws for their undying support.
\appendix{A}{Bosonic Dilaton Two-Point Function}
We present here the calculation for two dilatons on the sphere.
Reading off the  $\b{-1}{P}\b{1}{P}\b{-1}{Q}\b{1}{Q}$ contribution
from \bins (and with some help from {\it Mathematica}) gives
\eqn\fourform{\eqalign{\biggl(&-{h (r-\rr) \fp\gp\over 4 \rr |r| |\rr-r|}+
{\aa (1-f) h r (r-\rr) \fp\gp \over 4 m^2 |r| |\rr-r|} +\cr
&{ a \aa (1-f) r (r+\rr) \fp \gp \over 4 m^2 |r| |\rr-r|}+
{\aa (1-f) r^4 \rr^2 \fp \gp \over 4 m^2 n^2 |r| |\rr-r|}-\cr
&{\aa (1-f) g r^3 (r-\rr) \rr^2 \fp\gp \over 2 m^2 n^2 |r| |\rr-r|}+
{\aa (1-f) g^2 r \rr^2 (\rr-r)^2 (\rr+r) \fp \gp \over4 m^2 n^2 |r|
|\rr-r|}-\cr
& {a n \fp \gp \over 4 \rr m |r| |\rr-r|} +
{g (r-\rr) \rr \fp \gp \over 4 n |r| |\rr-r|}-\cr
&{\aa f \rr \fp\hp \over 4 m |\rr| |\rr-r|} + {f \gp \hp \over 4 |r|
|\rr|} +
{\aa (f-1) f r \rr \gp \hp \over 4 m^2 |r| |\rr|}\biggr)
     \b{-1}{P}\b{1}{P}\b{-1}{Q}\b{1}{Q}}}
This can be rewritten as
\eqn\derform{\eqalign{\biggl[
& \drrb\biggl({-h f\over\rr}\biggr)\drb\biggl({-g \over r}\biggr)+
\drrb\biggl({f g \rr\over n}\biggr)\drb\biggl({-g \over r}\biggr)+
\drrb\biggl({-a n f \over m \rr (\rr-r)}\biggr)
                              \drb\biggl({-g \over r}\biggr)+\cr
& \drrb\biggl({-a \aa (f-f^2/2)r (\rr+r) \over m^2(\rr-r)}\biggr)
                              \drb\biggl({-g \over r}\biggr)+
\drrb\biggl({\aa f(1-f)h r \over m^2}\biggr)
                              \drb\biggl({-g \over r}\biggr)+\cr
&\drrb\biggl({\aa f^2 h r \over 2 m^2}\biggr)
                              \drb\biggl({-g \over r}\biggr)+
 \drrb\biggl({-h  \over \rr}\biggr)
               \drb\biggl({-\aa f^2 \rr \over2 m (\rr-r)}\biggr)+\cr
&\drrb\biggl({\aa \rr^2 r (f-f^2/2) v\over  m^2 n^2 (\rr-r)}\biggr)
                              \drb\biggl({-g \over r}\biggr)\
              \biggr]\b{-1}{P}\b{1}{P}\b{-1}{Q}\b{1}{Q}}}
where $v=(mn-g r \rr (\rr-r))$.
The integral arising from these insertions can be re-written as the sum
of three pieces:
\eqn\dderi{
\int \drb\biggl({-g \over r}\biggr)
\dd \biggl[\biggl({a n f \over \rr m \y} +{h f\over \rr} -
                {g \rr f\over n} \biggr)\tfa\biggr]
     \b{-1}{P}\b{1}{P}\b{-1}{Q}\b{1}{Q}\ket{D}^P\otimes\ket{D}^Q,}
\eqn\dderii{\int\drrb\biggl({-h  \over \rr}\biggr)
 \dd\biggl[\biggl({\aa f^2 \rr \over2 m (\rr-r)}\biggr)\tfb\biggr]
     \b{-1}{P}\b{1}{P}\b{-1}{Q}\b{1}{Q}\ket{D}^P\otimes\ket{D}^Q,}
and
\eqn\dderiii{\eqalign{\int\drb\biggl({-g \over r}\biggr)
\dd\biggl[ \biggl(&{a \aa (f-f^2/2)r (\rr+r) \over m^2(\rr-r)}+
{-\aa f(1-f)h r \over m^2}+
 {-\aa f^2 h r \over 2 m^2}+\cr
&{-\aa \rr^2 r (f-f^2/2) v\over  m^2 n^2 (\rr-r)}\biggr)
\tfa\biggr]
     \b{-1}{P}\b{1}{P}\b{-1}{Q}\b{1}{Q}\ket{D}^P\otimes\ket{D}^Q.}}
The first term, \dderi, is the same as the contributions in \dtf that
arose in the dilaton-strong physical state contribution. Changing
variables to $y$ in favor of $\rr$ and integrating as before
just leaves
$$-2 \pi i (-2 a -2)\int\tf\drb\biggl({-g \over r}\biggr)
      \b{-1}{Q}\b{1}{Q}\ket{D}^Q$$
The remaining integral is just that in \onept for the one-point function
of the dilaton.
If this were the only contribution, then
we would have the dilaton equation with the proper normalization (cf.
\dspsi).
However, there are the terms in \dderii and \dderiii which are purely
contact terms and they spoil the dilaton equation. It is not clear why
these terms arise, but they are there and they agree with the local
calculation of Distler and Nelson \PRIV. The terms in \dderii and
\dderiii can be integrated as the other terms have been. In \dderii, we
eliminate $r$ in favor of $y$ and integrate (using the same contours as
in \conti) and find
$$  -2 \pi i \aa  \int\tff\drrb\biggl({-h  \over \rr}\biggr)
\b{-1}{P}\b{1}{P}\ket{D}^P$$
which is just the proper measure for the insertion of a dilaton at $P$
using the coordinates $\zeta_P$ in \zetaeq instead of at $\zeta_Q$ at$Q$.
This shouldn't be bothersome because this is a contact term that occurs
when $P$ and $Q$ coincide.
Finally, of the four terms in
\dderiii, the middle two integrate to 0 and the other two give
$$-2 \pi i (2 a\aa +\aa)\int\tf\drb\biggl({-g \over r}\biggr)
\b{-1}{Q}\b{1}{Q}\ket{D}^Q.$$ This is, of course, an insertion of the
dilaton at $Q$. Putting these three pieces together gives the dilaton
two-point function
$$\eqalign{\vev{DD}&=-2 \pi i (2(\aa-a+a \aa)-2)\vev{D}\cr
                            &=-2 \pi i (3/2-2)\vev{D}.}$$
which was the result stated in \dildil.
\appendix{B}{Heterotic Ghost Insertions}
We present here the ghost insertions that result from computing the
pushforwards of the tangent vectors associated to the moduli
corresponding to the locations of
the points $P$ and $Q$ by the coordinate family given in \slice. They were
computed with the help of {\it Mathematica.} The measure is formed by
multiplying these six contributions together (including delta functions
for the insertions for $\rho$ and $\tau$). In the text we only used the
contributions to the measure appropriate for the insertion of strong
physical states and the dilatons. The insertions are (keeping only the
terms with $R_1$ and $R_2$, and recalling that they are evaluated at
$r=\rb=0$)
\eqn\hetghost{\eqalign{
\B{\dbd{t}}=&
   -\biggl({2\over t}\biggr)\b{-1}{P}-
   \biggl({2\over t}+{4 p \over t}\biggr)\b{0}{P}-
   \biggl({2 p \over t}-{4 p^2\over t}+{a_2 |t| \fpt\over 2 t}\biggr)
        \b{1}{P}+\cr
  & \biggl({2 \tau\over t}+{4 p_1 \tau\over t}\biggr)\bet{\mha}{P}+
   \biggl({2 p \tau \over t}+{2 p_1 \tau \over t}-{12 p p_1 \tau\over t}
          +{a_1 \tau |t| \fpt\over t}\biggr)\bet{\ha}{P}-\cr
   &\biggl({ 2\over t}\biggr)\b{0}{Q}-
   \biggl({2 \pp \over t}+{\aa_2 |t| \fpt \over 2 t}\biggr)\b{1}{Q}+
   \biggl({2 \pp_1  \tau\over t}+{\aa_1\tau |t|\fpt\over t}\biggr)
   \bet{\ha}{Q}-\cr
  & \biggl({a_2 \tb \fpt \over 2 |t|}\biggr)\bb{1}{P}-
   \biggl({\aa_2 t \fpt \over 2 |t|}\biggr)\bb{1}{Q}+ \cdots
}}
$$\eqalign{
\cr
\B{\dbd{\tb}}=&
   -\biggl({a_2 t \fpt \over 2 |t|}\biggr)\b{1}{P}+
   \biggl({a_1 t \tau \fpt \over |t|}\biggr)\bet{\ha}{P}-
  \biggl({\aa_2 t \fpt \over 2 |t|}\biggr)\b{1}{Q}+\cr
 & \biggl({\aa_1 t \tau \fpt \over |t|}\biggr)\bet{\ha}{Q}-
   \biggl({2\over \tb}\biggr)\bb{-1}{P}-
   \biggl({2\over \tb}+{4 p\over\tb}\biggr)\bb{0}{P}-\cr
   &\biggl({2 p \over \tb}-{4 p^2 \over\tb}+{a_2 |t|\fpt\over 2 \tb}
                     \biggr)\bb{1}{P}-
   \biggl({2\over\tb}\biggr)\bb{0}{Q}-\cr
  &\biggl({2 \pp \over \tb}+{\aa_2|t| \fpt \over 2 \tb}
                 \biggr)\bb{1}{Q}+ \cdots
\cr
\B{\dbd{\tau}}=&
   -\bigl(\tau\bigr)\b{-1}{P}-
   \bigl({2 p \tau+2 p_1 \tau}\bigr)\b{0}{P}+
  \bigl({2 p^2 \tau -2 p p_1 \tau -p_1^2 \tau}\bigr)\b{1}{P}-\cr
  &\bigl(2\bigr)\bet{\mha}{P}-
   \bigl({2 p -2 p_1}\bigr)\bet{\ha}{P}-
   \bigl(\pp_1^2  \tau\bigr)\b{1}{Q}+
   \bigl(2 \pp_1 \bigr)\bet{\ha}{Q}+ \cdots
\cr
\B{\dbd{r}}=&
   -\bigl({1 \over t^2}\bigr)\b{-1}{P}-
   \biggl({2 p\over t^2}\biggr)\b{0}{P}+
   \biggl({2 p^2\over t^2}\biggr)\b{1}{P}+
   \biggl({2 p_1 \tau \over t^2}\biggr)\bet{\mha}{P}-
   \biggl({6 p p_1 \tau \over t^2}\biggr)\bet{\ha}{P}-\cr
  &\bigl({1\over t^2}\bigr)\b{-1}{Q}-
   \biggl({2 \pp \over t^2}\biggr)\b{0}{Q}+
   \biggl({2\pp^2 \over t^2}\biggr)\b{1}{Q}+
   \biggl({2 \pp_1 \tau \over t^2}\biggr)\bet{\mha}{Q}-
   \biggl({6 \pp \pp_1 \tau \over t^2}\biggr)\bet{\ha}{Q}+\cr
 &\bigl(R_2 \tb^2\bigr)\bb{-1}{P}+
  \bigl(2 R_2 (1-f) \tb^2 +2 R_2 p \tb^2\bigr)\bb{0}{P}+\cr
  & \bigl(R_2 (1-f)\tb^2 +2 R_2 (1+f)p \tb^2 -2 R_2 p^2 \tb^2
                 \bigr)\bb{1}{P}+
   \bigl(R_2 \tb^2\bigr)\bb{1}{Q}+\cdots
\cr
\B{\dbd{\rb}}=&
   \bigl(R_2 t^2 -2 R_1 t \rho \tau\bigr)\b{-1}{P}+\cr
  &\bigl(2 R_2 (1-f) t^2 +2 R_2 p t^2-2 R_1 (1-f) t \rho \tau -
    4 R_1 p t \rho \tau +\cr
   &\quad\quad2 R_1 p_1 t \rho \tau\bigr)\b{0}{P}+\cr
  &\bigl(R_2 (1-f) t^2 +2 R_2 (1+f) p t^2 -2 R_2 p^2 t^2 -2 R_1 (1-f) p t
\rho \tau+\cr
 &\quad\quad4 R_1 p^2 t\rho \tau+2 R_1 p_1 t \rho \tau -R_1 f p_1 t \rho
\tau -2 R_1 p p_1 t \rho \tau\bigr)\b{1}{P}-\cr
  &\bigl(2 R_1 t \rho+2 R_2 t^2 \tau +2 R_2 p_1 t^2 \tau\bigr)
                   \bet{\mha}{P}+\cr
  &\bigl(-2 R_1 t \rho +2 R_1 f t \rho -2 R_1 p t \rho -2 R_2 (1-f)t^2 \tau
    -2 R_2 p t^2 \tau -2 R_2 p_1 t^2 \tau-\cr
    &\quad4 R_2 f p_1 t^2 \tau+
    6 R_2 p p_1 t^2 \tau\bigr)\bet{\ha}{P}+
   \bigl(R_2 t^2 +2 R_1 \pp_1 t \rho \tau\bigr)\b{1}{Q}-
   \bigl(2 R_1 t \rho \tau\bigr)\bet{\ha}{Q}-\cr
  &\bigl({1\over\tb^2} \bigr)\bb{-1}{P}-
   \biggl({2 p \over \tb^2}\biggr)\bb{0}{P}+
   \biggl({2 p^2 \over \tb^2}\biggr)\bb{1}{P}-
   \bigl({1\over\tb^2}\bigr)\bb{-1}{Q}-
   \biggl({2 \pp \over \tb^2}\biggr)\bb{0}{Q}+\cr
  &\biggl({2 \pp\over\tb^2}\biggr)\bb{1}{Q}+\cdots
\cr
}$$\vfill\eject
$$\eqalign{
\B{\dbd{\rho}}=&
   \biggl(-{\rho\over t^2}+{2 \tau \over t}\biggr)\b{-1}{P}+
   \biggl(-{2 p \rho\over t^2}+{4 p \tau \over t}-{2 p_1 \tau \over t}
              \biggr)\b{0}{P}+\cr
  &\biggl({2 p^2 \rho \over t^2}-{4 p^2 \tau \over t}+{2 p p_1 \tau
        \over t}\biggr)\b{1}{P}+
   \biggl({2\over t}+{2 p_1 \rho \tau\over t^2}\biggr)\bet{\mha}{P}+\cr
  &\biggl({2 p \over t}-{6 p p_1 \rho\tau\over t^2}\biggr)\bet{\ha}{P}-
  \biggl({\rho \over t^2}\biggr)\b{-1}{Q}-
   \biggl({2 \pp \rho \over t^2}+{2 \pp_1 \tau \over t}
               \biggr)\b{0}{Q}+\cr
  &\biggl({2 \pp^2 \rho \over t^2}+{2 \pp \pp_1 \tau \over t }
                \biggr)\b{1}{Q}+
   \biggl({2\over t}+{2 \pp_1 \rho \tau \over t^2}
          \biggr)\bet{\mha}{Q}+\cr
  &\biggl({2 \pp \over t}-{6 \pp\pp_1\rho\tau\over t^2}
                     \biggr)\bet{\ha}{Q}+\cdots
}$$
where $p=a_2 (1-f)$, $p_1=a_1 (1-f)$, $\pp=\aa_2 (1-f)$, and
$\pp_1=\aa_1 (1-f)$. The dots represent higher terms ($ b_n, n\ge2$
$ \beta_n, n\ge3/2$) that do not contribute to the dilaton calculations
considered here.
\listrefs
\end